\documentclass[twocolumn,pra, aps,superscriptaddress,floatfix]{revtex4}

\usepackage{extarrows}
\usepackage{lineno}
\usepackage{mathptmx}
\usepackage{subfigure}
\usepackage{dcolumn}
\usepackage{amsmath,amssymb}
\usepackage{bm}
\usepackage{color}
\usepackage{overpic}
\usepackage{latexsym}
\usepackage{epstopdf}
\usepackage{color}
\usepackage[english]{babel}
\usepackage{latexsym}
\usepackage{stmaryrd}
\usepackage{bigints}
\usepackage{psfrag,graphicx} 
\usepackage{epsf} 
\usepackage{subfigure} 
\usepackage{amsmath} 
\usepackage{amssymb} 
\usepackage{amsfonts}
\usepackage{bm}
\usepackage{natbib}
\usepackage{epstopdf}
\usepackage{appendix}
\usepackage{relsize}

\definecolor{mygrey}{gray}{0.35}
\definecolor{myblue}{rgb}{0.2,0.2,0.8}
\definecolor{myzard}{cmyk}{0,0,0.05,0}
\definecolor{mywhite}{rgb}{1,1,1}
\definecolor{myred}{rgb}{1,0.,0.3}

\usepackage[colorlinks=true,citecolor=myblue,linkcolor=myred]{hyperref}
\def\be{\begin{equation}}
\def\ee{\end{equation}}
\def\ba{\begin{align}}
\def\enda{\end{align}}
\def\bi{\begin{itemize}}
\def\ei{\end{itemize}}

 \def\ee{\mathord{\rm e}}
 
 \def\ii{\mathord{\rm i}}

\def\half{\textstyle\frac{1}{2}}

\def\fourth{\textstyle\frac{1}{4}}

 \def\ee{\mathord{\rm e}}
 
 \def\ii{\mathord{\rm i}}

\def\half{\textstyle\frac{1}{2}}

\def\fourth{\textstyle\frac{1}{4}}

\renewcommand{\ii}{{\rm i}}
\renewcommand{\ee}{{\rm e}}

\def\beq{\begin{equation}}
\def\beq{\begin{equation}}
\def\eeq{\end{equation}}

 \newcommand{\ket}[1]{|#1\rangle}
 \newcommand{\bra}[1]{\langle #1|}

\begin{document}


\title[Short Title]{Renormalization group flows  for Wilson-Hubbard matter and the topological Hamiltonian}

\author{E. Tirrito} 
\affiliation{ICFO - Institut de Ciencies Fotoniques, The Barcelona Institute of Science and Technology, Av. Carl Friedrich Gauss 3, 08860 Castelldefels (Barcelona), Spain} 
\author{M. Rizzi}
\affiliation{Johannes Gutenberg-Universit\"{a}t, Institut f\"{u}r Physik,  Staudingerweg 7, 55099 Mainz, Germany}
\author{G. Sierra}
\affiliation{Instituto de F\'{i}sica Te\'{o}rica UAM/CSIC, Universidad Aut\'{o}noma de Madrid, Madrid, Spain}
\author{M. Lewenstein} 
\affiliation{ICFO - Institut de Ciencies Fotoniques, The Barcelona Institute of Science and Technology, Av. Carl Friedrich Gauss 3, 08860 Castelldefels (Barcelona), Spain}
\affiliation{ICREA, Pg. Lluís Companys 23, 08010 Barcelona, Spain}
\author{A. Bermudez}
\affiliation{Departamento de F\'{i}sica Te\'{o}rica, Universidad Complutense, 28040 Madrid, Spain}

\begin{abstract}
Understanding the robustness of topological phases of matter in the presence of
interactions poses a difficult challenge in modern 
condensed matter,  showing  interesting connections to high energy physics.
In this work, we  leverage  these connections to present a complete  analysis of the continuum long-wavelength 
description of a generic class of correlated topological insulators: Wilson-Hubbard topological matter. 
We show that a Wilsonian renormalization group (RG) approach,  combined with the so-called topological Hamiltonian, provide   a quantitative route to understand interaction-induced topological phase transitions that occur in Wilson-Hubbard matter. We benchmark  two-loop RG predictions for a quasi-1D Wilson-Hubbard  model by means of exhaustive numerical simulations based on matrix product states (MPS). The agreement of the RG predictions with MPS simulations motivates the extension of the RG calculations to higher-dimensional 
topological insulators.  

\end{abstract}

\maketitle
\setcounter{tocdepth}{2}
\begingroup
\hypersetup{linkcolor=black}
\tableofcontents
\endgroup

\section{Introduction}
\label{sec:introduction}

Condensed-matter and high-energy physics study phenomena at vastly different scales and, yet,  these can often  be understood by the same  concepts and unifying principles.  One of the major examples of the cross-fertilization of fundamental ideas between  these two disciplines is the theory of spontaneous symmetry breaking, which is both paramount  to our current understanding of  phases and phase transitions in condensed matter~\cite{landau,landau_ginzburg,anderson}, and  to the standard model describing  elementary  particles and their interactions~\cite{nambu, goldstone,  higgs}. These parallelisms become clearer under the light of a  common language: the theory of interacting quantum fields~\cite{fradkin_book,qft_book}. In this context, again,  a multi-disciplinar approach based on concepts of scaling and critical phenomena leads to the so-called renormalization group~\cite{wk_rg}, which has turned out to be the key to understand many-body effects in condensed-matter systems~\cite{shankar_rg}, or the very definition of a relativistic quantum field theory~\cite{rg_conceptual}.

Beyond these fruitful connections, there is also the possibility of finding condensed-matter analogues of models initially introduced in the realm of high-energy physics, or vice versa. In this way, one may not only use a common framework to understand widely different phenomena, but actually observe the same phenomenon in two different scenarios. Graphene is a representative and well-known example of this situation~\cite{graphene_review}, as the low-energy electronic excitations of this semi-metal act as analogues of a relativistic quantum field theory of massless Dirac fermions. Another recent example of this trend lies in the physics of topological insulators, which are insulating phases of matter that are not characterized by local order parameters but, instead, by certain topological invariants~\cite{Bernevig}. Some of these topological insulators can be represented by  minimal models that are condensed-matter analogues of relativistic quantum field theories of massive Wilson fermions~\cite{wilson_fermions}, which originally appeared in the context of lattice gauge theories for elementary particle physics~\cite{lattice_book}.

A question of current interest in the field of topological insulators~\cite{ti_interactions_reviews} and symmetry-protected topological (SPT) phases~\cite{spt_review} is to explore  correlation effects, such as the possibility of finding interaction-induced  quantum phase transitions that separate these SPT phases from other non-topological ground-states. Let us note that, although these topological quantum phase transitions cannot be understood from the  principle of spontaneous symmetry breaking mentioned above, they can still be characterized by the closure of a many-body energy gap, such that scaling phenomena should also be relevant.
Accordingly, the connection of continuum relativistic quantum field theories (QFTs), scaling, and the renormalization group (RG) to these topological phases might offer a systematic route to understand  correlation effects. In this paper, we carefully explore this possibility for certain types of correlated topological insulators: {\it Wilson-Hubbard topological matter}, which can be described in terms of a relativistic QFT of massive Wilson fermions with four-Fermi interactions in the vicinity of a topological band inversion point. We show that a Wilsonian RG offers a neat qualitative picture to understand these topological phases, and a quantitative scheme to obtain the phase boundaries that separate them from other non-topological phases.
 
This article is organized as follows. In Sec.~\ref{sec:func_int}, we start by reviewing the use of relativistic QFTs of massive Wilson fermions as long-wavelength representatives of topological insulators~\ref{subsec:wilson_qft_spt}. We then include interactions leading to Wilson-Hubbard topological matter, and  discuss the Euclidean functional integral that contains all relevant information about the possible phases and phase transitions~\ref{subsec:euc_action_wilson_hubbard}. In~\ref{subsec:RG_generic}, we start by reviewing the Wilsonian  approach to the RG of interacting QFTs. We then move on to  discuss generic features of the RG flows for  Wilson-Hubbard matter at tree level, which offer a neat connection between the long-wavelength limit and the flat-band limit of topological insulators. We also discuss the use of the RG flows of the QFT parameters to obtain interaction-induced corrections to the so-called topological Hamiltonian~\cite{self_energy_topological,topological_hamiltonian},  providing a straightforward route to calculate the corresponding  many-body topological invariants. In Sec.~\ref{sec:RG_1D_SPT}, we apply the  RG to a one-dimensional model of Wilson-Hubbard matter: the imbalanced Creutz-Hubbard ladder~\ref{subsec:CH_ladder}. This model of an interacting topological insulator yields a perfect playground to test the qualitative and quantitative validity of the RG approach beyond tree level, as one may compare two-loop RG corrections to quasi-exact numerical simulations based on matrix product states~\ref{subsec:loop_expansion_ch}. In~\ref{subsec:entanglement_spect}, we provide a further benchmark of the RG calculations and the renormalized topological Hamiltonian by characterizing the bi-partite correlations at the topological quantum phase transition. Finally, in Sec.~\ref{sec:conclusions}, we present our conclusions and outlook.
 
\section{Functional integrals   for topological insulators with Hubbard-type interactions}
\label{sec:func_int}

\subsection{Quantum field theories as representatives of non-interacting topological insulators}
\label{subsec:wilson_qft_spt}

In this section, we describe  a generic lattice Hamiltonian~\cite{TI_field_theories,10_fold_ryu,wilson_atoms} that can host different classes of non-interacting topological insulators~\cite{review_top_insulators,Bernevig}, as well as a continuum quantum field theory that captures its long-wavelength properties, including the underlying topological features.
This will serve us to review certain aspects of topological insulators, and to set the notation used throughout this manuscript.  
 
 \vspace{1ex}
{\it (a) Continuum Dirac QFTs and topological invariants.--}
 \vspace{0.5ex}
 
 \noindent
A fundamental ingredient of this work is the { massive Dirac quantum field theory} in a $D=(d+1)$-dimensional Minkowski space-time~\cite{qft_book}, which  can serve as a building block to construct a  representative QFT for various  topological insulating phases~\cite{10_fold_ryu}. In the Hamiltonian formulation, this quantum field theory (QFT)  can be written as
\beq
\label{eq:dirac_qft}
H_{\rm D}\!=\!\bigintssss_{{\Lambda}^d}\!\frac{{\rm d}^dk}{(2\pi)^d}\Psi_\mu^{\dagger}(\boldsymbol{k})[h_{\rm D}(\boldsymbol{k})]^{\mu,\nu}\Psi_{\nu}^{\phantom{\dagger}}(\boldsymbol{k}),\hspace{1ex}h_{\rm D}(\boldsymbol{k})=\alpha_i\hspace{0.15ex}k^i+m\beta,
\eeq
where $\Psi^{\phantom{\dagger}}\!\!\!\!(\boldsymbol{k}), \Psi^{{\dagger}}\!(\boldsymbol{k})$ are the spinor fermionic field operators fulfilling $\{\Psi_\mu^{\phantom{\dagger}}(\boldsymbol{k}), \Psi_{\nu}^{{\dagger}}(\boldsymbol{k}')\}=(2\pi)^d\delta_{\mu,\nu}\delta^{(d)}(\boldsymbol{k}-\boldsymbol{k}')$, and the momentum   lies below a certain ultra-violet (UV)  cutoff to regulate the QFT, namely $\boldsymbol{k}\in {\Lambda}^d$ if $|k_i|\leq \Lambda_{\rm c}$ (i.e.  $\Lambda^d\to\mathbb{R}^d$ as the cutoff is removed $\Lambda_{\rm c}\to\infty$). In the above expression, $\{\alpha_i\}_{i=1}^d$ and  $\beta$ are the so-called Dirac matrices, which are mutually anti-commuting Hermitian matrices that square to the identity, and we use natural units   $\hbar=c=1$ together with   Einstein's summation criterion. Depending on the particular choice of the Dirac matrices~\cite{rqm_book}, the single-particle Hamiltonian $h_{\rm D}(\boldsymbol{k})$ may, or may not, respect  the following discrete symmetries:   time-reversal symmetry   $\mathsf{T}$ is fulfilled if $U_{\mathsf{T}}h_{\rm D}(-\boldsymbol{k})^*U_{\mathsf{T}}^\dagger=h_{\rm D}(\boldsymbol{k})$; particle-hole   symmetry $\mathsf{C}$ takes place when $U_{\mathsf{C}}h_{\rm D}(-\boldsymbol{k})^*U_{\mathsf{C}}^\dagger=-h_{\rm D}(\boldsymbol{k})$; and  the so-called sub-lattice   symmetry $\mathsf{S}$ occurs if $U_{\mathsf{S}}h_{\rm D}(\boldsymbol{k})U_{\mathsf{S}}^\dagger=-h_{\rm D}(\boldsymbol{k})$, where we have introduced various unitary rotations $U_{\mathsf{T}},U_{\mathsf{C}},U_{\mathsf{S}}$~\cite{table_top_insulators}.  We note that these discrete symmetries can be related  to the ten-fold classification of symmetric spaces~\cite{az_ten_fold} via the
corresponding time-evolution operator $U_{\rm D}(t)=\ee^{-\ii t h_{\rm D}(\boldsymbol{k})}$.

To give concrete examples   that shall be used throughout this work, let us consider {\it (i)} the $(2+1)$-dimensional QFT~\eqref{eq:dirac_qft} with $\alpha_1=\sigma^x$, $\alpha_2=\sigma^y$, and $\beta=\sigma^z$ expressed in terms of  Pauli matrices. In this case, there is no  unitary operator that can fulfill any of the above transformations, and the Dirac QFT thus explicitly breaks the time-reversal, particle-hole, and sub-lattice symmetries. In this case, the time-evolution operator $U_{\rm D}(t)=\ee^{-\ii t h_{\rm D}(\boldsymbol{k})}$ lies in the so-called unitary  symmetric space~\cite{10_fold_ryu}, labeled as $\mathsf{A}$. From this parent Hamiltonian, one may use a Kaluza-Klein-type dimensional reduction by compactifying the second spatial direction into a vanishingly small circle~\cite{TI_field_theories,10_fold_ryu}.   {\it (ii)} The resulting $d=1$ Dirac QFT~\eqref{eq:dirac_qft} with $\alpha_1=\sigma^x$,  and $\beta=\sigma^z$, now respects the sub-lattice symmetry with $U_{\mathsf{S}}=\sigma^y$, but breaks both time-reversal and particle-hole symmetries. In this case, the time-evolution operator belongs to the chiral unitary symmetric space, labeled as $\mathsf{AIII}$. We note that this idea of dimensional reduction allows one to find representatives of all ten possible symmetric spaces starting from a  higher-dimensional Dirac QFT~\cite{TI_field_theories,10_fold_ryu}.

To understand the use of these QFTs as building blocks of topological insulators,  let us consider the zero-temperature groundstate  $\ket{\epsilon_{\rm gs}}=\prod_{\boldsymbol{k}\in\Lambda^d}\ket{\epsilon_-(\boldsymbol{k})}$, which is obtained by occupying all the single-particle modes  $\ket{\epsilon_-(\boldsymbol{k})}$ with   energies $\epsilon_-(\boldsymbol{k})=-\sqrt{m^2+\boldsymbol{k}^2}$. 
Depending on the dimensionality and the discrete symmetries  of the Hamiltonian introduced above, one can define a {\it topological invariant} that characterizes each ground-state. These topological  invariants, which cannot be modified by perturbations that respect the corresponding  symmetries unless a quantum phase transition takes place, can be defined through momentum integrals of either the so-called {\it Chern characters} or  {\it Chern-Simons forms}~\cite{10_fold_ryu}. For the examples cited above, one finds  that: {\it (i)}  the $d=2$ Dirac QFT in the $\mathsf{A}$ class is characterized by the integral of the first Chern character $\mathsf{ch}_1=\frac{\ii}{2\pi}\partial_{k^i}\bra{\epsilon_-(\boldsymbol{k})}\partial_{k^j}\ket{\epsilon_-(\boldsymbol{k})}dk^i\times dk^j$, 
\beq
\label{eq:chern_ch}
\mathsf{ Ch}_1=\int_{\Lambda^2}\mathsf{ ch}_1=\frac{m}{2|m|}.
\eeq
This topological invariant, which is known as the first Chern number, cannot change unless $m=0$, which signals  a quantum phase transition. {\it (ii)} For the $d=1$ Dirac QFT in the $\mathsf{AIII}$ class, the integral of the first Chern-Simons form $\mathsf{Q}_1=\frac{\ii}{2\pi}\bra{\epsilon_-({k})}\partial_{k}\ket{\epsilon_-({k})}dk$ yields the  Chern-Simmons invariant
\beq
\label{eq:chern_form}
\mathsf{ CS}_1=\int_{\Lambda^1}\mathsf{ Q}_1=\frac{m}{4|m|}.
\eeq
In contrast to the Chern number, this quantity is not gauge invariant, and one typically defines an associated  Wilson loop 
 $\mathsf{W}_1=\ee^{\ii 2\pi \mathsf{ CS}_1}$, which gives a quantized topological invariant that cannot change unless $m=0$. We emphasize  that similar topological invariants can be constructed  for any particular higher-dimensional Dirac QFT~\cite{10_fold_ryu}, which allows one to find the required building blocks  for all 10 possible topological insulators-superconductors~\cite{table_top_insulators}, excluding the appearance of additional crystalline symmetries.

\vspace{1ex}
{\it (b) Discretized QFTs, fermion doubling and invariants.--}
 \vspace{0.5ex}
 
 \noindent
From a condensed-matter perspective~\cite{fradkin_book}, the above Dirac QFT~\eqref{eq:dirac_qft} will arise as a  low-energy approximation that tries to capture the relevant physics  of a particular material at long wavelengths $\xi\gg a$ or, equivalently, at low momenta $|{k_i}|\leq \Lambda_{\rm c}\ll 2\pi/a$. Here, $a$ is the lattice constant of the material, which serves as natural UV regulator of the QFT. Paradigmatic examples of  massless  Dirac QFTs in condensed-matter setups appear in  the so-called one-dimensional Luttinger liquids~\cite{LL_review} and two-dimensional graphene~\cite{graphene_review}. On the other hand, for a non-vanishing mass/gap, these continuum QFTs~\eqref{eq:dirac_qft} can serve as  building blocks to construct  a   low-energy approximation of topological insulators. Let us now introduce some peculiarities of the lattice regularization from the perspective of lattice field theory (LFT)~\cite{lattice_book}.

A naive approach  is to discretize the spatial derivatives of the Dirac Hamiltonian that lead to the linear dependence on momentum in Eq.~\eqref{eq:dirac_qft}. By introducing a  Bravais lattice   $ \Lambda_{\ell}=a\mathbb{Z}^d=\{\boldsymbol{ x}:x_{i}/a\in\mathbb{Z},\forall i=1,\cdots,d\}$, the corresponding  Hamiltonian LFT can be expressed as follows
\beq
\label{eq:naive_DF_lattice}
H_{\rm D}=\!\!\sum_{\boldsymbol{ x}\in\Lambda_\ell}\!\!a^d\!\left(\!\Psi^{\dagger}(\boldsymbol{ x})\frac{\ii \alpha_i}{2a} \Psi\!\left(\boldsymbol{ x}+a\boldsymbol{u}^i\right)+m\,\Psi^{\dagger}(\boldsymbol{ x})\frac{\beta}{2} \Psi(\boldsymbol{ x})+{\rm H.c.} \!\right)\!,
\eeq
where $\{\boldsymbol{u}^i\}_{i=1}^d$ are the  unit vectors of the Bravais lattice. Moreover, the   lattice spinor fields $\Psi(\boldsymbol{x}), \Psi^{{\dagger}}\!(\boldsymbol{x})$  display the desired  anti-commutation algebra in the continuum limit   $\{\Psi_\mu^{\phantom{\dagger}}(\boldsymbol{x}), \Psi_{\nu}^{{\dagger}}(\boldsymbol{x}')\}=\frac{1}{a^d}\delta_{\mu,\nu}\delta_{\boldsymbol{x},\boldsymbol{x}'}\to\delta_{\mu,\nu}\delta^{(d)}(\boldsymbol{x}-\boldsymbol{x}')$. Transforming the field operators  to momentum space, 
\beq
\Psi_\mu^{\phantom{\dagger}}(\boldsymbol{x})=\int_{{\rm BZ}^d}\!\frac{{\rm d}^dk}{(2\pi)^d}\ee^{\ii\boldsymbol{k}\cdot\boldsymbol{x}}\hspace{0.25ex}\Psi_\mu^{\phantom{\dagger}}(\boldsymbol{k}),
\eeq
one obtains a QFT  similar to Eq.~\eqref{eq:dirac_qft}, where momenta now lie within the so-called Brillouin zone $\Lambda^d\to{\rm BZ}^d=(-\frac{\pi}{a},\frac{\pi}{a}]^{\times^d}$, and the single-particle Hamiltonian is
\beq
h_{\rm D}(\boldsymbol{k})=\frac{1}{a} \alpha_i\sin (k^ia)+m\beta.
\eeq

The energy spectrum of this lattice Hamiltonian  becomes $\epsilon_{\pm}(\boldsymbol{k})\approx\pm\sqrt{m^2+\boldsymbol{k}^2}$ at long wavelengths $\xi\sim k_i^{-1}\gg a$, and thus reproduces the energy of the massive Dirac fermion. However, there are other points at the borders of the Brillouin zone that lead to a similar dispersion relation (i.e. Dirac points), and thus give rise to additional relativistic fermions, the so-called {\it fermion doublers}. In fact, there is an even number  $N_{\rm D}=2^d$ of Dirac points, labelled by $\boldsymbol{ k}_{\boldsymbol{ n}}=\frac{\pi}{a} n_i\boldsymbol{u}^i$ with $n_i\in\{0,1\}$,  which lead to a dispersion relation that is approximately described by a massive relativistic particle $\epsilon_{\pm}(\boldsymbol{ k}_{\boldsymbol{ n}}+\boldsymbol{k})\approx\pm\sqrt{m^2+\boldsymbol{k}^2}$ at long wavelengths $\xi\sim k_i^{-1}\gg a$. The effective QFT around each of these points corresponds to an  instance of the massive Dirac QFT $h_{\rm D}(\boldsymbol{ k}_{\boldsymbol{ n}}+\boldsymbol{k})\approx\alpha_i^{\boldsymbol{n}}\hspace{0.15ex}k^i+m\beta$ with a different choice of the Dirac matrices $\alpha_i^{\boldsymbol{n}}=(-1)^{n_i}\alpha_i$. Let us note that this effective QFT is defined within a certain long-wavelength cutoff where the linearization is valid, such that $\Lambda^d=(-\pi/l_{\rm c},\pi/l_{\rm c}]^{\times^d}$ in  Eq.~\eqref{eq:dirac_qft}. As will become  clear  below,  using  the perspective of the renormalization group (RG)~\cite{wk_rg}, we shall approach this continuum limit by setting the parameters of the lattice   Hamiltonian close to a quantum phase transition, where the characteristic correlation length diverges $\xi/l_{\rm c}\to \infty$, and  effectively ${\Lambda}^d\to\mathbb{R}^d$. Accordingly, we get $N_{\rm D}$ copies of the desired continuum QFT~\eqref{eq:dirac_qft}.

From the perspective of  LFT, the presence of  fermion doublers with $\boldsymbol{n}\neq 0$ is an unfortunate nuisance, as they  will inevitably couple to the target $\boldsymbol{n}= 0$ Dirac QFT as soon as additional interaction terms are included~\cite{lattice_book}. Moreover, the presence of the doublers is a generic feature for different lattice discretizations, and is related to the difficulty of incorporating chiral symmetry on the lattice~\cite{nn_theorem}. 

From the perspective of topological insulators, the presence of these doublers is already important at the non-interacting level. In the lattice, the even number of relativistic fermions can be divided into two sets $S_{\pm}$ with opossite chiralities, each of which contains an   equal number  $|S_{\pm}|=2^{(d-1)}$of Dirac points: $\boldsymbol{n}\in S_{\pm}$ if  $(-1)^{\sum_i n_i}=\pm 1$. This sign difference is translated into an opposite Chern character~\eqref{eq:chern_ch} or Chern-Simons form~\eqref{eq:chern_form}, such that the topological invariant obtained by integrating over the whole Brillouin zone  vanishes (i.e. summing an equal number of  positive and negative contributions over all  Dirac points yields a zero topological invariant).  Accordingly,  due to the presence of the spurious fermion doublers, this na{i}ve lattice Hamiltonian~\eqref{eq:naive_DF_lattice} fails at reproducing the  non-vanishing topological invariant in Eq.~\eqref{eq:chern_ch} or~\eqref{eq:chern_form} of the single massive Dirac QFT~\eqref{eq:dirac_qft}. For the representative examples discussed above, one finds that: {\it (i)} there are a $N_{\rm D}=4$ Dirac points for the $d=2$ Dirac QFT in the $\mathsf{A}$ class, such that the integral of the first Chern character $
\mathsf{ Ch}_1=\int_{\Lambda^2}\mathsf{ ch}_1=\frac{m}{2|m|}(1-1-1+1)=0$. {\it (ii)} For the $d=1$ Dirac QFT in the $\mathsf{AIII}$ class, there are $N_{\rm D}=2$ Dirac points, such that  integral of the first Chern-Simons form
$
\mathsf{ CS}_1=\int_{\Lambda^1}\mathsf{ Q}_1=\frac{m}{4|m|}(1-1)=0$, leading to a trivial Wilson loop $\mathsf{W}_1=1$. We conclude that the groundstate of such  naive lattice models corresponds to a trivial  insulator, and not to the desired SPT phase.

\vspace{1ex}
{\it (c) Wilson  and continuum  QFTs  of  topological insulators.--}
 \vspace{0.5ex}
 
 \noindent
We now discuss a possible route to get around the effect of the spurious doublers, which is well-known in LFT~\cite{domain_wall,dw_fermions_various}, and corresponds to a generic model of non-interacting topological insulators~\cite{TI_field_theories,10_fold_ryu,wilson_atoms}. The route is based on Wilson's prescription to construct lattice Hamiltonians that yield a different mass $m_{\boldsymbol{n}}$ for each of the  fermion doublers~\cite{wilson_fermions}. This can be achieved by modifying Eq.~\eqref{eq:naive_DF_lattice}, and  introducing   the so-called  {\it Wilson-fermion  Hamiltonian} LFT
\beq
\label{eq:Wilson_lattice}
H_{\rm W}=H_{\rm D}+\sum_{\boldsymbol{ x}\in\Lambda_\ell}\!\!a^d\!\left(\Psi^{\dagger}(\boldsymbol{ x})\frac{ \delta m_i \beta}{2} \Psi\!\left(\boldsymbol{ x}+a\boldsymbol{u}^i\right)+{\rm H.c.} \!\right)\!,
\eeq
where the parameters $\delta m_i$ quantify certain  mass shifts. In this case, transforming the lattice Hamiltonian to momentum space, $
H_{\rm W}\!=\!\bigintssss_{{\rm BZ}^d}\!\frac{{\rm d}^dk}{(2\pi)^d}\Psi_\mu^{\dagger}(\boldsymbol{k})[h_{\rm W}(\boldsymbol{k})]^{\mu,\nu}\Psi_{\nu}^{\phantom{\dagger}}(\boldsymbol{k})$, yields the following single-particle Hamiltonian
\beq
h_{\rm W}(\boldsymbol{k})=\frac{1}{a} \alpha_i\sin (k^ia)+(m+\delta m_i\cos(k^ia))\beta.
\eeq
We perform again a long-wavelength approximation around the different momenta $\boldsymbol{ k}_{\boldsymbol{ n}}=\frac{\pi}{a} n_i\boldsymbol{u}^i$, namely setting $\boldsymbol{k}\to\boldsymbol{k}+\boldsymbol{ k}_{\boldsymbol{ n}}$ such that   $|\boldsymbol{k}|\ll \pi/l_{\rm c}$. The Wilson-fermion LFT~\eqref{eq:Wilson_lattice} yields  a {\it continuum Wilson-fermion QFT} described by $N_{\rm D}$ instances of the massive Dirac QFT~\eqref{eq:dirac_qft}, each describing a relativistic fermion with a different Wilson mass 
\beq
\label{eq:wilson_qft}
H_{\rm W}\approx\bigintssss_{{\Lambda}^d}\frac{{\rm d}^dk}{(2\pi)^d}\sum_{\boldsymbol{n}}\Psi_{\boldsymbol{n},\mu}^{\dagger}(\boldsymbol{k})[h_{\rm D}^{\boldsymbol{n}}(\boldsymbol{k})]^{\mu,\nu}\Psi_{\boldsymbol{n},\nu}^{\phantom{\dagger}}(\boldsymbol{k}).
\eeq
Here, we have introduced the spinorial fermionic operators  $\Psi^{{\dagger}}_{\boldsymbol{n}}(\boldsymbol{k}), \Psi^{\phantom{\dagger}}_{\boldsymbol{n}}(\boldsymbol{k})$ that  create-annihilate Dirac fermions    with  a Wilson mass $m_{\boldsymbol{n}}$, and the single-particle Hamiltonians  
\beq
\label{eq:eq:wilson_single_particle}
h_{\rm D}^{\boldsymbol{n}}(\boldsymbol{k})=\alpha^{\boldsymbol{n}}_i\hspace{0.15ex}k^i+m_{\boldsymbol{n}}\beta, \hspace{2ex}m_{\boldsymbol{n}}=m+ \sum_i(-1)^{n_i}\delta m_i.
\eeq
We note once more that this effective QFT is defined within a certain cutoff, such that ${\Lambda}^d=(-\pi/l_{\rm c},\pi/l_{\rm c}]^{\times^d}\to\mathbb{R}^d$ in the vicinity of a  critical point (i.e. continuum limit $\xi/l_{\rm c}\to\infty$). This limit, as well as the meaning of the approximation symbol in Eq.~\eqref{eq:wilson_qft}, will be discussed below in light of the RG.

From a QFT perspective,  setting  the lattice parameters $m$ and $\delta m_i$ in such a way  that   $m_{\boldsymbol{n}}\sim a^{-1}\gg m_{\boldsymbol{0}}\geq0$ $\forall\, \boldsymbol{n}\neq \boldsymbol{0}$, effectively sends all fermion doublers to very high energies (on the order of the UV cutoff). Accordingly,   it could be expected that they will not contribute to the low-energy phenomena described by the single  Dirac fermion at  $\boldsymbol{ k}_{\boldsymbol{ 0}}=\boldsymbol{0}$. For $\delta m_i=-m/d$, it is only the  Dirac point  at $\boldsymbol{ k}_{\boldsymbol{ 0}}$ that is  massless, and   can serve as the starting point to study quarks coupled to gauge fields in lattice  quantum chromodynamics~\cite{wilson_fermions,lattice_book}. 

From the  perspective of topological insulators, the situation is complicated by the fact that the topological invariants are non-local quantities obtained by integrating over all momenta~\eqref{eq:chern_ch}-\eqref{eq:chern_form}, and are thus also sensitive to the doublers even if they lie at very large energies. In fact, the presence of the doublers is  crucial  to turn the above  invariants in Eq.~\eqref{eq:chern_ch} or~\eqref{eq:chern_form} into {\it integer topological numbers}, making the Wilson-fermion QFT~\eqref{eq:wilson_qft}  a  generic quantum-field-theory representative of topological insulators~\cite{10_fold_ryu}. 

This can be  easily observed  for the examples introduced above: {\it (i)}  the $d=2$ Wilson-fermion  QFT representative  of the $\mathsf{A}$ class of topological insulators  can be related to   the integer quantum Hall effect~\cite{iqhe}, where  integer Chern numbers  underlie the  quantization of the transverse conductance~\cite{chern_tknn}.  For the corresponding groundstate of the Wilson QFT~\eqref{eq:wilson_qft},  the integral of the first Chern character yields 
\beq
\label{eq:ch_1_wilson}
\mathsf{ Ch}_1=\int_{\Lambda^2}\mathsf{ ch}_1=\sum_{\boldsymbol{n}}{p}_{\boldsymbol{n}}\frac{m_{\boldsymbol{n}}}{2|m_{\boldsymbol{n}}|}=\sum_{\boldsymbol{n}}\half{p}_{\boldsymbol{n}}{\rm sgn}(m_{\boldsymbol{n}}),
\eeq
where we have introduced ${p}_{\boldsymbol{n}}={\rm exp}\{\ii\pi\sum_in_i\}$. Accordingly, if  the lattice parameters $\delta m_i$ are such that   a {\it mass inversion} occurs for some of the Wilson fermions, it becomes possible to obtain a non-vanishing integer-valued topological invariant $
\mathsf{ Ch}_1\in\mathbb{Z}$, which can be related to the plateaus of the quantum Hall effect observed at integer fillings~\cite{chern_tknn}. In addition, the edge states responsible for the transverse conductance in the integer quantum Hall effect~\cite{iqhe_edges} have a counterpart in the Wilson LFT: they correspond to the so-called domain-wall fermions~\cite{domain_wall,dw_fermions_various}, which are masless Dirac fermions localized at the boundaries of the  lattice.

Something similar occurs for the {\it (ii)} $d=1$ Wilson  LFT   of the $\mathsf{AIII}$ class of topological insulators, where 
\beq
\label{eq:cs_1_wilson}
\mathsf{ CS}_1=\int_{\Lambda^1}\mathsf{ Q}_1=\sum_{\boldsymbol{n}}p_{\boldsymbol{n}}\frac{m_{\boldsymbol{n}}}{4|m_{\boldsymbol{n}}|}=\sum_{\boldsymbol{n}}\fourth{p}_{\boldsymbol{n}}{\rm sgn}(m_{\boldsymbol{n}}).
\eeq
One thus obtains  a non-trivial Wilson loop $\mathsf{W}_1=\ee^{\ii 2\pi \mathsf{ CS}_1}=- 1$  whenever an odd number of mass inversions take place. Once again, if one introduces an interface between the topological phase and a non-topological one (e.g. considering a finite system with boundaries), gapless  edge excitations  localized to the interface appear, which are related to the zero-energy Jackiw-Rebbi modes localised at a mass-inversion point~\cite{Jackiw_Rebbi}.

Let us note that alternative discretizations that also lead to Dirac fermions with different Wilson masses, thus displaying  topological features and edge states,   have also appeared in  condensed-matter  contexts~\cite{fradkin,haldane,disorder_iqhe,kane_mele}. From the perspective  of LFTs, these edge modes correspond to lower-dimensional versions of the aforementioned domain-wall fermions. From the perspective of topological insulators, the properties of these phases can be shown to be robust against any perturbation, e.g. disorder,  that respects the symmetry of the given symmetry class. A  problem of current interest in the community is to understand the interplay of these topological properties and correlation effects brought by  interactions~\cite{ti_interactions_reviews}.

\subsection{Euclidean action and functional integrals for correlated topological insulators}
\label{subsec:euc_action_wilson_hubbard}

In this section, we present a functional-integral description of  the previous generic lattice Hamiltonian~\eqref{eq:Wilson_lattice} when Hubbard-type interactions~\cite{hubbard_model} between the fermions are included. This functional integral will be the starting point to develop RG techniques that  allow us to understand the fate of the topological phases as the interactions are switched on.

Let us consider the Wilson lattice Hamiltonian~\eqref{eq:Wilson_lattice} in the so-called lattice units $a=1$, and introduce Hubbard-type  contact interactions describing the repulsion of fermions. This leads to the following {\it Wilson-Hubbard lattice} Hamiltonian
\beq
\label{eq:wilson_hubbard_ham}
H_{\rm WH}= H_{\rm W}+\sum_{\nu,\mu}\sum_{\boldsymbol{ x}\in\Lambda_\ell}\Psi^{\dagger}_\mu(\boldsymbol{ x})\Psi_\nu^{\dagger}(\boldsymbol{ x})\frac{ u_{\mu\nu}}{2} \Psi_\nu(\boldsymbol{ x})\Psi_\mu(\boldsymbol{ x}),
\eeq
where we have introduced  interaction strengths $u_{\mu\nu}>0$. In general, the relation of these parameters to the spinorial indexes will depend on the nature of the orbitals that determine the fermionic spinors $\Psi(\boldsymbol{ x}), \Psi^\dagger(\boldsymbol{ x})$. For the previous examples of  $\mathsf{A}$ and  $\mathsf{AIII}$ topological insulators, due to Pauli exclusion principle, one simply finds $u_{\mu\nu}=(1-\delta_{\mu,\nu})u$ for $u>0$.

We are interested in computing the partition function of  the Wilson-Hubbard Hamiltonian $\mathsf{Z}={\rm Tr}\{{\rm exp}(-\beta H_{\rm WH})\}$,  where $\beta$ is the inverse temperature, as it contains all the relevant information about the possible quantum phase transitions in the zero-temperature limit $\beta\to\infty$. This partition function can be expressed as a functional integral by means of  fermionic coherent states~\cite{fradkin_book}. We thus  introduce  Grassmann spinors  $\psi(\boldsymbol{x},\tau), \overline{\psi}(\boldsymbol{x},\tau)$ at each lattice point $\boldsymbol{x}\in\Lambda_{\ell}$ and   imaginary time $\tau\in(0,\beta)$, which are composed of mutually anti-commuting Grassman fields  $\{\psi_\mu(\boldsymbol{x},\tau), \overline{\psi}_\nu(\boldsymbol{x}',\tau')\}=\{\psi_\mu(\boldsymbol{x},\tau), {\psi}_\nu(\boldsymbol{x}',\tau')\}=\{\overline{\psi}_\mu(\boldsymbol{x},\tau), \overline{\psi}_\nu(\boldsymbol{x}',\tau')\}=0$. Since the Wilson-Hubbard Hamiltonian is already normal-ordered, one can readily express the zero-temperature partition function as a functional integral $\mathsf{Z}=\int[{\rm d}\overline{\psi}{\rm d}{\psi}]\ee^{-S_{\rm WH}[\bar{\psi},{\psi}]}$, where the Euclidean action is a functional of the  Grassmann fields 
\beq
S_{\rm WH}[\overline{\psi},{\psi}]=\int_0^\infty\!\!{\rm d}\tau\!\sum_{\boldsymbol{x}\in\Lambda_\ell}\!\big(\overline{\psi}(\boldsymbol{x},\tau)\partial_\tau{\psi}(\boldsymbol{x},\tau)+H_{\rm WH}(\overline{\psi},{\psi})\big),
\eeq
and $H_{\rm WH}(\overline{\psi},{\psi})$ results from substituting the fermion field operators  by  Grassmann variables  in the normal-ordered Hamiltonian~\eqref{eq:wilson_hubbard_ham}.
By Fourier transforming to frequency and momentum space  $\psi_\mu^{\phantom{\dagger}}(\boldsymbol{x},\tau)=\int_{\mathbb{R}}\frac{{\rm d}\omega}{2\pi}\int_{{\rm BZ}^d}\!\frac{{\rm d}^dk}{(2\pi)^d}\ee^{\ii(\boldsymbol{k}\cdot\boldsymbol{x}-\omega\tau)}\hspace{0.25ex}\psi_\mu^{\phantom{\dagger}}(\boldsymbol{k},\omega)
$,  the action becomes 
\beq
\label{eq:Wilson_hubbard_action}
S_{\rm WH}[\overline{\psi},{\psi}]=S_{\rm W}[\overline{\psi},{\psi}]+\delta S_{\rm W}[\overline{\psi},{\psi}]+\delta S_{\rm H}[\overline{\psi},{\psi}].
\eeq
 Here, we have introduced the free action $S_{\rm W}$ for the  Wilson QFT~\eqref{eq:wilson_qft} with single-particle Hamiltonian~\eqref{eq:eq:wilson_single_particle}, namely
\beq
\label{eq:long_wavelength_action}
S_{\rm W}=\bigintssss_{\boldsymbol{k},\omega}\sum_{\boldsymbol{n}}\overline{\psi}_{\boldsymbol{n},\mu}(\boldsymbol{k},\omega)[-\ii\omega\mathbb{I}+h_{\rm D}^{\boldsymbol{n}}(\boldsymbol{k})]^{\mu,\nu}\psi_{\boldsymbol{n},\nu}^{\phantom{\dagger}}(\boldsymbol{k},\omega),
\eeq
where the integration symbol is 
$\int_{\boldsymbol{k},\omega}=\int_{\mathbb{R}}\frac{{\rm d}\omega}{2\pi}\int_{{\Lambda}^d}\!\!\frac{{\rm d}^dk}{(2\pi)^d}$. Since this QFT arises at  long wavelengths, we must also include   corrections  to this approximation in $\delta S_{\rm W}$, namely
\beq
\label{eq:short_wavelength_action}
\delta S_{\rm W}=\bigintssss_{\boldsymbol{k},\omega}\sum_{\boldsymbol{n}}\overline{\psi}_{\boldsymbol{n},\mu}(\boldsymbol{k},\omega)[\delta h_{\rm D}^{\boldsymbol{n}}(\boldsymbol{k})]^{\mu,\nu}\psi_{\boldsymbol{n},\nu}^{\phantom{\dagger}}(\boldsymbol{k},\omega),
\eeq
where we have introduced
\beq
\label{eq:single_particle_short_wavelength}
\delta h_{\rm D}^{\boldsymbol{n}}(\boldsymbol{k})=(-1)^{n_i} \left[ \left( \sin (k^i) - k^i \right) \alpha_i + \delta m_i \left( \cos (k^i) - 1 \right) \beta \right].
\eeq
Finally, the Hubbard interactions lead to the quartic term
\beq
\label{eq:S_int}
\delta S_{\rm H}=\bigintsss_{\{\boldsymbol{k},\omega\}}\!\sum_{\{\boldsymbol{n}\}}\frac{u_{\mu\nu}}{2}\overline{\psi}_{\boldsymbol{n}_4,\mu}(4)\overline{\psi}_{\boldsymbol{n}_3,\nu}(3)\psi_{\boldsymbol{n}_2,\nu}^{\phantom{\dagger}}(2)\psi_{\boldsymbol{n}_1,\mu}^{\phantom{\dagger}}(1),
\eeq
where we have introduced the short-hand notations   $\psi_{\boldsymbol{n}_j,\mu}^{\phantom{\dagger}}(j)=\psi_{\boldsymbol{n}_j,\mu}^{\phantom{\dagger}}(\boldsymbol{k}_j,\omega_j)$ for $j\in\{1,2,3,4\}$~\cite{shankar_rg}, $\sum_{\{\boldsymbol{n}\}}=\sum_{\boldsymbol{n}_1,\boldsymbol{n}_2,\boldsymbol{n}_3,\boldsymbol{n}_4}$, and   $\int_{\{\boldsymbol{k},\omega\}}=\prod_{j}\int_{\boldsymbol{k}_j,\omega_j}(2\pi)^2\delta(\omega_4+\omega_3-\omega_2-\omega_1)\delta(\boldsymbol{q}_4+\boldsymbol{q}_3-\boldsymbol{q}_2-\boldsymbol{q}_1)$, where $\boldsymbol{q}_j=\boldsymbol{k}_{\boldsymbol{n}_j}+\boldsymbol{k}_{j}$ is the momentum about the different Dirac points. The goal of this work is to study the interplay of the quadratic and quartic terms in the Wilson-Hubbard action~\eqref{eq:Wilson_hubbard_action} from the point of view of the RG~\cite{shankar_rg}.

\subsection{Renormalization group  flows of Wilson fermions and the topological Hamiltonian}
\label{subsec:RG_generic}
In this section, we discuss generic properties of interacting topological insulators in light of the RG. Since we are interested in the continuum QFT description of interacting topological insulators, we must focus on phenomena at long length scales $\xi$,  where it makes sense to partition  the action~\eqref{eq:Wilson_hubbard_action} into a long-wavelength term $S_{\rm W}$~\eqref{eq:long_wavelength_action}, and the perturbations caused by  shorter-wavelength corrections $\delta S_{\rm W}$~\eqref{eq:short_wavelength_action} and by the interactions $\delta S_{\rm H}$~\eqref{eq:S_int}. Since $\xi\gg l_{\rm c}$, the physical properties of   interest should not depend on our choice of the cutoff in the action~\eqref{eq:Wilson_hubbard_action}. Therefore, we should be able to change the cutoff $l_{\rm c}\to l_{\rm c}'=s l_{\rm c}>l_{\rm c}$ for $s>1$ (i.e. $\Lambda_{\rm c}\to \Lambda_{\rm c}'=\Lambda_{\rm c}/s<\Lambda_{\rm c}$) without modifying the physics. This can only be achieved if one allows for the microscopic coupling parameters of the Wilson-Hubbard action~\eqref{eq:Wilson_hubbard_action} to run with the cutoff $\{g_i(\Lambda_{\rm c})\}$  (i.e. $\{m_{\boldsymbol{n}}(\Lambda_{\rm c}), u_{\mu\nu}(\Lambda_{\rm c})\})$, which is the essence of the renormalization group~\cite{shankar_rg, rg_conceptual}.
The Wilsonian RG prescription allows us to calculate such an {\it RG flow} in a systematic fashion, and associates the change of the bare couplings to the dressing of the  long-wavelength fields by the short-wavelength modes that must be integrated out as  the cutoff is lowered~\cite{wk_rg}.

\vspace{1ex}
{\it (a) Wilsonian RG by coarse graining and rescaling.--}
 \vspace{0.5ex}
 
 \noindent
By writing the Grassmann fields in terms of the short-wavelength (fast) and long-wavelength (slow) modes,  $\psi_{\boldsymbol{n},\mu}^{\rm f}(\boldsymbol{k},\omega)=\psi_{\boldsymbol{n},\mu}(\boldsymbol{k},\omega)$ for $\Lambda_{\rm c}\geq|\boldsymbol{k}|\geq\Lambda_{\rm c}/s$, and $\psi_{\boldsymbol{n},\mu}^{\rm s}(\boldsymbol{k},\omega)=\psi_{\boldsymbol{n},\mu}(\boldsymbol{k},\omega)$ for $\Lambda_{\rm c}/s\geq|\boldsymbol{k}|\geq0$, respectively, the Wilsonian RG can be divided into two steps: 
 
{\it (i)} In the first one, one {\it coarse grains} the action by integrating out the fast modes, such that the new  coarse-grained action can be expressed, up to an irrelevant constant, as follows
\beq
\label{eq:effective_action}
S'_{\rm WH}[\overline{\psi}^{\rm s},{\psi}^{\rm s}]=S_{\rm W}[\overline{\psi}^{\rm s},{\psi}^{\rm s}]-\log\left\langle\ee^{-\delta S_{\rm WH}\big[\overline{\psi}^{\rm s},\overline{\psi}^{\rm f},{\psi}^{\rm s},{\psi}^{\rm f}\big]}\right\rangle_{\!\!{\rm f}},
\eeq
In this expression, we have grouped   all the perturbations in  $\delta S_{\rm WH}[\overline{\psi}^{\rm s},\overline{\psi}^{\rm f},{\psi}^{\rm s},{\psi}^{\rm f}]=\delta S_{\rm W}[\overline{\psi}^{\rm f},{\psi}^{\rm f}]+\delta S_{\rm H}[\overline{\psi}^{\rm s},\overline{\psi}^{\rm f},{\psi}^{\rm s},{\psi}^{\rm f}]$, and defined their expectation value with respect to the non-interacting partition function   $\langle \ee^{-\delta S_{\rm WH}[\overline{\psi}^{\rm s},\overline{\psi}^{\rm f},{\psi}^{\rm s},{\psi}^{\rm f}]}\rangle_{{\rm f}}=\int_{d\Lambda}\ee^{-S_{\rm W}[\overline{\psi}^{\rm f},{\psi}^{\rm f}]}\ee^{-\delta S_{\rm WH}[\overline{\psi}^{\rm s},\overline{\psi}^{\rm f},{\psi}^{\rm s},{\psi}^{\rm f}]}/\int_{d\Lambda}\ee^{-S_{\rm W}[\overline{\psi}^{\rm f},{\psi}^{\rm f}]}$, where we have  introduced a short-hand notation for the integral over the fast modes $\int_{d\Lambda}=\int_{\Lambda_{\rm c}/s\leq |\boldsymbol{k}|\leq\Lambda_{\rm c}}\prod_k{\rm d}\overline{\psi}^{\rm f}(\boldsymbol{k},\omega){\rm d}{\psi}^{\rm f}(\boldsymbol{k},\omega)$. To evaluate this expectation value, we will make use of the free-fermion propagator (i.e.  the single-particle Green's function)
\beq
\label{eq:green_function_free}
G^0_{\mu\nu}(\ii\omega, \boldsymbol{k})=\langle\overline{\psi}_{{\mu}}(\boldsymbol{k},\omega)\psi_{\nu}(\boldsymbol{k},\omega)\rangle_{\rm f}=\left[\big(\ii\omega- h_{\rm D}^{\boldsymbol{n}} (\boldsymbol{k})\big)^{-1}\right]_{\mu\nu}.
\eeq

{\it (ii)} In the second step of the RG transformation, one  {\it rescales} the momentum  and frequency 
\beq
\label{eq:rescaling}
\boldsymbol{k}\to\boldsymbol{k}'=s\boldsymbol{k}, \hspace{1ex}\omega\to\omega'=s\omega,
\eeq
such that the original cutoff is restored $\Lambda_{\rm c}'\to s\Lambda_{\rm c}'=\Lambda_{\rm c}$.  Note that frequencies and wave-vectors are rescaled equally, which can be traced back to  a dynamical exponent  $z=1$, and the emerging  Lorentz invariance in  the continuum limit~\eqref{eq:wilson_qft}.
In this situation,  one can compare the original~\eqref{eq:Wilson_hubbard_action} and coarse-grained~\eqref{eq:effective_action} actions, trying to extract the dressing of the couplings $\{m_{\boldsymbol{n}}(\Lambda_{\rm c}), u_{\mu\nu}(\Lambda_{\rm c})\}$ by the fast modes that have been integrated out during the   coarse-graining step. This rescaling must be accompanied by   a {\it scale transformation} for the slow modes 
\beq
\psi^{\rm s}_{\boldsymbol{n},\mu}(\boldsymbol{k},\omega)\to{\psi}'_{\boldsymbol{n},\mu}(\boldsymbol{k}',\omega')=s^{-\Delta_{\psi}}{\psi}^{\rm s}_{\boldsymbol{n},\mu}(\boldsymbol{k}'/s,\omega'/s),
\eeq
where $\Delta_{\psi}$ is  the so-called scaling dimension of the fermion fields, which are  homogeneous under the rescaling of momenta and frequencies. 

The scaling dimension is chosen in such a way that the massless part of the free action~\eqref{eq:long_wavelength_action}  becomes  invariant under  rescaling, i.e. the terms of the QFT~\eqref{eq:wilson_qft}-\eqref{eq:eq:wilson_single_particle} for vanishing couplings $m=\delta m_i=u_{\mu\nu}=0$, generally denoted as $\boldsymbol{g}=\boldsymbol{0}$, do not get modified.  Let us  note that it is customary in RG treatments of QFT  to define dimensionless couplings $\boldsymbol{g}$~\cite{rg_conceptual}. 

Accordingly, the action for the naive Dirac fermions corresponds to a {\it fixed point} under the RG subsequent coarse-graining and rescaling transformations, as it remains unaltered
 $
S_{\rm WH}[\overline{\psi},\psi; \boldsymbol{g}=\boldsymbol{0},\Lambda_{\rm c}]\xlongrightarrow{\text{\it (i)}} S_{\rm WH}'[\overline{\psi}^{\rm s},\psi^{\rm s}; \boldsymbol{g}=\boldsymbol{0},\Lambda_{\rm c}']
\xlongrightarrow{\text{\it (ii)}}  S_{\rm WH}'[\overline{\psi}',\psi'; \boldsymbol{g}=\boldsymbol{0},\Lambda_{\rm c}]=S_{\rm WH}[\overline{\psi},\psi;\boldsymbol{g}=\boldsymbol{0},\Lambda_{\rm c}]$. This is achieved by choosing the fermion scaling dimension of $\Delta_{\psi}=(d+2)/2$ as a function of the spatial dimensionality $d$~\cite{shankar_rg}. 
In the present case, considering that  $\Delta_{\psi}=(d+2)/2$, and that  the action~\eqref{eq:Wilson_hubbard_action} is  dimensionless, one finds that the interactions $u_{\mu\nu}$ must be   dimensionless, whereas the   mass can be adimensionalized  by $m_{\boldsymbol{n}}\to m_{\boldsymbol{n}}\Lambda_{\rm c}$  after substituting $m\to m\Lambda_{\rm c}$ and $\delta m_i\to\delta m_i\Lambda_{\rm c}$.

To proceed with the RG program in practice, and extract the flow of the couplings, we must be able to calculate the coarse-grained action~\eqref{eq:effective_action}. In  perturbative RG, this is performed by applying a perturbative expansion in some small parameter. The cumulant expansion offers a systematic procedure to reach  the desired order of the perturbation
\begin{multline}
\label{eq:cum_exp_action}
\log \left\langle\ee^{-\delta S_{\rm WH}\big[\overline{\psi}^{\rm s},\overline{\psi}^{\rm f},{\psi}^{\rm s},{\psi}^{\rm f}\big]}\right\rangle_{\!\!{\rm f}}= \\
=\sum_{n=1}^{\infty}\frac{(-1)^n}{n!}\left\langle \delta S_{\rm WH}\big[\overline{\psi}^{\rm s},\overline{\psi}^{\rm f},{\psi}^{\rm s},{\psi}^{\rm f}\big]^n\right\rangle_{\rm c,f}
\end{multline}
where $\langle \delta S_{\rm WH}[\overline{\psi}^{\rm s},\overline{\psi}^{\rm f},{\psi}^{\rm s},{\psi}^{\rm f}]^n\rangle_{\rm c,f}$ is the $n$-th cumulant of the perturbation obtained by  integrating over the fast modes, e.g. the mean $\langle \delta S_{\rm WH}^1\rangle_{\rm c,f}=\langle \delta S_{\rm WH}\rangle_{\rm f}$, the variance $\langle \delta S_{\rm WH}^2\rangle_{\rm c,f}=\langle \delta S_{\rm WH}^2\rangle_{\rm f}-\langle \delta S_{\rm WH}\rangle_{\rm f}^2$, and so on.  

\vspace{1ex}
{\it (b) Tree-level RG flow and adiabatic band flattening--}
 \vspace{0.5ex}
 
 \noindent
With this machinery, we can now discuss some generic  features of the RG for the Wilson-Hubbard matter~\eqref{eq:Wilson_hubbard_action},  making connections to the underlying topological insulators, and to the possible quantum phase transitions that connect them to other non-topological states of matter. At zero order of the cumulant expansion, it is straightforward to calculate the RG flow of the dimensionless Wilson masses $m_{\boldsymbol{n}}$~\eqref{eq:eq:wilson_single_particle} appearing in the long-wavelength action~\eqref{eq:long_wavelength_action}. According to Eq.~\eqref{eq:effective_action}, after the two RG steps, the masses   only get a contribution from rescaling
\beq
\label{eq:beta_mass}
m_{\boldsymbol{n}}\Lambda_{\rm c}\to  m_{\boldsymbol{n}}\Lambda_{\rm c} \, s, \hspace{3ex}\beta_{m_{\boldsymbol{n}}}=\frac{{\rm d}m_{\boldsymbol{n}}}{{\rm d}\log s} =m_{\boldsymbol{n}},
\eeq
where $\beta_{m_{\boldsymbol{n}}}$ are the beta functions generally defined as $\beta(\boldsymbol{g})=\partial \boldsymbol{g} / \partial \log s $ where $s$ represents the energy scale.

Accordingly, the effect of the Wilson masses gets amplified ($s>1$) as one integrates more and more short-wavelength modes. Therefore, one says that the mass terms are {\it relevant perturbations} that take us away from the infrared (IR) RG fixed point of the  $N_{\rm D}$ massless  non-interacting Dirac fermions. Note also that the above RG flow respects the sign of the Wilson masses at the initial cutoff $\Lambda_{\rm c}$. Hence,  if ${\rm sgn}(m_{\boldsymbol{n}})=\pm 1$ at $\Lambda_{\rm c}$, then  $m_{\boldsymbol{n}}\to\pm \infty$  in the IR limit (i.e. positive masses become more positive, whereas negative masses become more negative, as one focuses on longer and longer length scales). This will be of crucial importance for our study of  the topological features of the Wilson-Hubbard model.

It is also  straightforward to calculate the flow of the shorter-wavelength perturbations~\eqref{eq:short_wavelength_action} at first-order of the cumulant expansion. The RG transformation affects the terms in Eq.~\eqref{eq:single_particle_short_wavelength} as follows
$
\delta h_{\rm D}^{\boldsymbol{n}}(\boldsymbol{k})\xlongrightarrow{\text{\it (i)+(ii)}} s\delta h_{\rm D}^{\boldsymbol{n}}(\boldsymbol{k}/s)=\sum_{\ell=1}^{\infty} \frac{(-1)^{\ell+n_i}}{(2\ell+1)!}\alpha_i(k^i)^{2\ell+1}s^{-2\ell}+\frac{(-1)^{\ell+n_i}}{(2\ell)!}\delta m_i\Lambda_{\rm c}(k^i)^{2\ell}\beta s^{1-2\ell}$, all of which decrease under the RG  since $s>1$, and $\ell\geq 1$. 

These results  allow us to understand the approximation symbol in Eqs.~\eqref{eq:wilson_qft}  formally as the result of the RG flow towards the long-wavelength IR limit (i.e. the  shorter-wavelength corrections are {\it irrelevant perturbations} in the RG sense, and can be thus safely discarded). Moreover, in the non-interacting regime, the RG offers an interesting picture of the topological insulating phases based solely on the action~\eqref{eq:long_wavelength_action}. The IR limit corresponds to sending the Wilson masses to $m_{\boldsymbol{n}}\to\pm \infty$ such that, in comparison, the dispersion of the bands becomes  vanishingly small  and we can approximate $\epsilon^{\boldsymbol{n}}_\pm(k)\approx\pm m_{\boldsymbol{n}}$. Therefore, the RG transformation amounts to a continuous deformation  into an equivalent  flat-band  model  $h_{\rm D}^{\boldsymbol{n}}(k)\xlongrightarrow{\text{\it (i)+(ii)}}|m_{\boldsymbol{n}}|\sum_{\boldsymbol{k}\in\Lambda^d}(P_{+}(\boldsymbol{k})-P_{-}(\boldsymbol{k}))$, where $P_{\pm}(\boldsymbol{k})$ are orthogonal projectors onto the flat bands~\cite{Bernevig}. Since the signs of the Wilson masses ${\rm sgn}(m_{\boldsymbol{n}})$ are preserved under the RG flow, the running of the Wilson masses can be considered as an adiabatic transformation that preserves the above  topological invariants~\eqref{eq:ch_1_wilson}-\eqref{eq:cs_1_wilson}. Accordingly, if there is a mass inversion at the original cutoff responsible for  a non-vanishing integer topological invariant, the groundstate of the model at the IR limit will indeed correspond to a topological insulating state of the equivalent   flat-band  model. 

Despite the fact that the RG flow of the Wilson masses seems to imply that the excitations  become infinitely heavy in the IR, and that the effective low-energy QFT should be trivial~\cite{rg_conceptual}, non-zero integer topological invariants~\eqref{eq:ch_1_wilson}-\eqref{eq:cs_1_wilson} indicate that the system can indeed display a non-trivial quantized response at low energies. In fact, even if the bulk excitations become infinitely heavy, the  Wilson-fermion LFT with mass inversion would display massless edge excitations  in a finite lattice with boundaries, such that the IR behavior is not trivial. The conservation of the topological invariant under the RG is a different manifestation of the non-triviality of the IR QFT with boundaries (i.e. bulk-edge correspondence). The quantum phase transition between a topological insulator and a trivial band insulator will thus be marked by the mass inversion for   one  of the  $N_{\rm D}$  Wilson fermions, which we label as $\boldsymbol{n}_\star$. Hence,  the critical point of the non-interacting model is marked by $m_{\boldsymbol{n}_\star}=0$, and thus corresponds to the RG fixed point of a  massless Dirac fermion, which controls the scaling properties of the quantum phase transition. 

\vspace{1ex}
{\it (c) Interacting RG  and the topological Hamiltonian.--}
 \vspace{0.5ex}
 
 \noindent
The question now is to study how this neat RG picture is modified as the quartic Hubbard interactions are switched on. Considering again the first-order cumulant $\langle \delta S_{\rm H}[\overline{\psi}^{\rm s},\overline{\psi}^{\rm f},{\psi}^{\rm s},{\psi}^{\rm f}]\rangle_{\rm f}$ at the so-called {\it tree level} (i.e. processes that only involve the slow Grassmann fields), the Hubbard interaction strengths flow with
\beq
\label{eq:hubb_tree_level}
u_{\mu\nu}\to s^{1-d}u_{\mu\nu}, \hspace{3ex}\beta_{u_{\mu\nu}}=\frac{{\rm d}u_{\mu\nu}}{{\rm d}\log s} =(1-d)u_{\mu\nu}.
\eeq
which implies that the interactions are a {\it marginal perturbation} for the $d=1$ case, and irrelevant in higher dimensions. 

Let us note that, even when the interactions are irrelevant, they can modify how the Wilson masses run, and affect considerably the shape of the phase diagram. 
Our   goal then is to go beyond  tree level, and explore how the interactions change the above RG flow and the topological insulating phases. On general grounds, we expect that the interactions will modify the $\beta_{m_{\boldsymbol{n}}}$ function of the  masses~\eqref{eq:beta_mass}, such that the bare Wilson masses get renormalized by the Hubbard couplings 
\beq
\label{dressed_masses}
m_{\boldsymbol{n}}\to \tilde{m}_{\boldsymbol{n}}=m_{\boldsymbol{n}}+\delta m_{\boldsymbol{n}}(u_{\mu\nu}).
\eeq 
More generally, as a consequence of the interaction,  the free-fermion propagator~\eqref{eq:green_function_free} of the long-wavelength action~\eqref{eq:long_wavelength_action} will be transformed into $G^0(\ii\omega, \boldsymbol{k})\to G(\ii\omega,\boldsymbol{k})$, where 
\beq
G^{-1}(\ii\omega, \boldsymbol{k})=\ii\omega- h_{\rm D}^{\boldsymbol{n}}(\boldsymbol{k})-\Sigma_{\rm D}^{\boldsymbol{n}}(\ii\omega,\boldsymbol{k}),
\eeq
and $\Sigma_{\rm D}^{\boldsymbol{n}}(\ii\omega, \boldsymbol{k})$ is the so-called self-energy, which includes various many-body scattering events that modify the propagation of fermions. This quantity includes, among other non-static effects, the aforementioned renormalization of the fermion masses. Regarding its connection to topological insulators, it has recently been demonstrated that only the static part of the self-energy $\Sigma_{\rm D}^{\boldsymbol{n}}(0,\boldsymbol{k})$ carries the relevant information for the topological properties~\cite{self_energy_topological}. Accordingly, one can define a {\it topological Hamiltonian}~\cite{topological_hamiltonian} that incorporates the effects of interactions on the topological properties as follows
\beq
\label{top_ham}
h_{\rm top}^{\boldsymbol{n}}(\boldsymbol{k})=h_{\rm D}^{\boldsymbol{n}}(\boldsymbol{k})+\Sigma_{\rm D}^{\boldsymbol{n}}(0,\boldsymbol{k}).
\eeq
The calculation of topological invariants for interacting systems then parallels the discussion of the previous sections, as the topological Hamiltonian can be interpreted as a dressed single-particle Hamiltonian. This notion has become very useful in the literature, as numerical tools such as dynamical mean-field theory~\cite{topological_hamiltonian_dmft_coupled_cluster} or quantum Monte Carlo~\cite{topological_hamiltonian_qmc} are ideally suited to calculate the zero-frequency self energy.
 
In our work, we shall be interested in understanding how the analytical techniques based on perturbative RG can be combined with the topological Hamiltonian to understand the fate of topological insulators as interactions are increased. From the preceding discussion, it is clear that the aforementioned renormalization of the Wilson masses will enter the zero-frequency self-energy as $\Sigma_{\rm D}^{\boldsymbol{n}}(0,\boldsymbol{k})=\delta m_{\boldsymbol{n}}(u_{\mu\nu})\beta$, where $\beta$ corresponds here to the Dirac matrix. Using the topological Hamiltonian~\eqref{top_ham}  provides us with a direct way to extend the topological invariants~\eqref{eq:ch_1_wilson}-\eqref{eq:cs_1_wilson} to the correlated regime, as we can calculate exactly the previous topological invariants. In our case, we simply need to consider the renormalized masses 
\beq
\mathsf{ Ch}_1=\sum_{\boldsymbol{n}}\half{p}_{\boldsymbol{n}}{\rm sgn}(\tilde{m}_{\boldsymbol{n}})
\eeq
 for the two-dimensional $\mathsf{A}$ Chern insulators in~\eqref{eq:ch_1_wilson},  and 
 \beq
 \label{eq:CS_AIII_interacting}
 \mathsf{ CS}_1=\sum_{\boldsymbol{n}}\fourth{p}_{\boldsymbol{n}}{\rm sgn}(\tilde{m}_{\boldsymbol{n}})
 \eeq
  for the one-dimensional $\mathsf{AIII}$ topological insulators in~\eqref{eq:cs_1_wilson}.

 A quantitative calculation of the renormalization of the Wilson masses  will depend on the particular model under study. In the following, we shall develop this RG program in detail for a one-dimensional case, which serves as a neat playground where our analytical predictions can be confronted to  precise numerical simulations~\cite{dmrg,creutz_hubbard}. We will use this example  to test  qualitatively  the above  RG picture of interacting topological insulators. In contrast to the numerical or analytical techniques used  in~\cite{creutz_hubbard}, which are not immediately available for higher spatial dimensions, the RG scheme beyond tree-level can  be directly  extended   to higher-dimensional models, which we leave for future detailed  studies.

\section{Renormalization group  flows  and the topological Hamiltonian of  $\textsf{AIII}$ topological insulators}
\label{sec:RG_1D_SPT}

\subsection{The imbalanced Creutz-Hubbard ladder}
\label{subsec:CH_ladder}

In this section, we consider a   simple modification~\cite{creutz_hubbard} of a lattice  model  leading to a $d=1$ Wilson-fermion Hamiltonian~\cite{creutz_ladder}. This model will be used as a testbed for the RG  of interaction effects in   $\textsf{AIII}$ topological insulators.  

\vspace{1ex}
{\it (a) Wilson-Hubbard matter on a $\pi$-flux two-leg ladder.--}
 \vspace{0.5ex}
 
 \noindent
The { imbalanced Creutz model} consists of spinless fermions  on a two-leg ladder~\cite{creutz_hubbard}. These fermions   are created and annihilated by  $c_{j,\ell}^{\dagger},c_{j,\ell}^{\phantom{\dagger}}$, where    $j\in\{1,\dots,N\}$ labels the lattice sites within the upper or lower legs   $\ell\in\{\rm u,d\}$ of the ladder, and evolve according to the tight-binding Hamiltonian
\beq
\label{eq:cl}
H_{\rm C}=\sum_{j,\ell}\left(-t_{\ell}c_{j+1,\ell}^{\dagger}c_{j,\ell}^{\phantom{\dagger}}-t_{\rm x}c_{j+1,\ell}^{\dagger}c_{j,\bar{\ell}}^{\phantom{\dagger}}+\frac{\Delta\epsilon_\ell}{4} c_{j,\ell}^{\dagger}c_{j,\ell}^{\phantom{\dagger}}+{\rm H.c.}\right).
\eeq
Here, $t_\ell= t\ee^{-\ii \pi s_\ell/2}$ represents the horizontal hopping strength dressed by a magnetic $\pi$-flux, $t_{\rm x}$ stands for the diagonal hopping,  $\Delta\epsilon_{\ell}=s_\ell\Delta\epsilon$ with $\Delta\epsilon>0$ is an energy imbalance between the legs of the ladder, and we use the notation  $s_{\rm u}= 1$ ($s_{\rm d}= -1$), and 
$\bar{\ell}= {\rm d} (\bar{\ell}= {\rm u}) $ for $\ell= {\rm u}$ ($\ell= {\rm d}$). In this section, we  start by setting $\hbar=1$, and derive  an effective microscopic effective speed of light $c$ with which one can normalise all parameters, achieving the desired natural units.

In the thermodynamic limit,  the rungs of the ladder play the role of the 1D Bravais lattice $ja\to\boldsymbol{x}\in\Lambda_\ell$ introduced above Eq.~\eqref{eq:naive_DF_lattice}, while the ladder index $\ell\in\{\rm u, d\}$ plays the role of the spinor degrees of freedom of the Fermi field $\Psi(\boldsymbol{x})=(c_{j,{\rm u}}, c_{j,{\rm d}})^{\rm t}$.  Making a  gauge transformation $c_{j,{\ell}}\to \ee^{\ii\pi j/2}c_{j,{\ell}}$, one finds that the above imbalanced Creutz model can be rewritten as a 1D Wilson  Hamiltonian~\eqref{eq:Wilson_lattice} in lattice units $a=1$,  provided that one makes the following identification of Dirac matrices $\alpha_1=\sigma^x$,  $\beta=\sigma^z$, and bare dimensionless parameters $m=\Delta\epsilon/4t_{\rm x}$ $\delta m=t/t_{\rm x}$. According to the general discussion above, we should find $N_{\rm D}=2$ Wilson fermions with masses $
m_{n}\propto \big(\Delta\epsilon\pm 4t\big)$,
 such that the  critical  point separating  topological and normal insulators corresponds to $\Delta\epsilon=4t$~\cite{creutz_hubbard}. 

Since these effects should be independent of the gauge choice, we will stick to the original lattice formulation~\eqref{eq:cl}, which essentially implies that the Dirac points will be shifted from ${k}_{{n}}=\pi n\to{k}_{{n}}=\frac{\pi}{2}(1-2 n)$, where we recall that $n\in\{0,1\}$ (i.e. ${k}_{{0}}=\pi/2$ and ${k}_{{1}}=-\pi/2$). This can be readily seen in  momentum-space, where
$
H_{\rm C}=\int_{-\pi}^{+\pi}\frac{dk}{2\pi}\Psi^\dagger\!(k) h_{\rm C}(k)\Psi(k)
$ with the single-particle Hamiltonian 
\beq
\label{eq:single_part_creutz}
h_{\rm C}({k})=-2t_{\rm x}\sigma^x\cos k+\left(\half\Delta\epsilon+2t\sin k\right)\sigma^z.
\eeq
The energy dispersion around the above Dirac points $k\to{ k}_{{ n}}+{k}$, becomes $\epsilon_{\pm}({ k}_{{ n}}+{k})\approx\pm\sqrt{\tilde{m}_n^2c^4+c^2{k}^2}$, where the effective speed of light is $c=2t_{\rm x}$ and the Wilson masses are $ \tilde{m}_{n}=\frac{1}{4t_{\rm x}^2}\left(\frac{\Delta\epsilon}{2}+(-1)^n 2t\right)$. To make contact with the previous convention $c=1$, we normalize the Hamiltonian by the effective speed of light, and thus obtain the dimensionless masses 
\beq
\label{eq:wilson_mass_1d}
{m}_{n}=\left(\frac{\Delta\epsilon}{4t_{\rm x}}+(-1)^n\frac{t}{t_{\rm x}} \right),
\eeq
which confirm the above expectation $
m_{n}\propto \big(\Delta\epsilon\pm 4t\big)$. 

In this non-interacting regime, one finds that the integral of the Chern-Simon form $\mathsf{Q}_1=\frac{\ii}{2\pi}\bra{\epsilon_-({k})}\partial_{k}\ket{\epsilon_-({k})}dk$ in the chiral basis over the whole Brillouin zone, which is proportional to the so-called Zak's phase~\cite{zak_phase},  yields $\mathsf{CS}_1=\int_{-\pi}^{\pi}\mathsf{Q}_1=\half\theta(1-\Delta\epsilon/4t)$, where $\theta(x)$ is the Heaviside step function. This expression is fully equivalent to   Eq.~\eqref{eq:cs_1_wilson}, which was obtained in the long-wavelength limit by adding the contributions of the pair of Wilson fermions
\beq
\label{eq:top_inv_CH}
\mathsf{ CS}_1=\fourth\big({\rm sgn}(m_{0})-{\rm sgn}(m_{1})\big).
\eeq
Let us note that  $\Delta\epsilon<4t$ corresponds to the inverted mass regime, since   $m_{1}\propto\left(\Delta\epsilon- 4t\right)<0$, and $m_{0}\propto\left(\Delta\epsilon+ 4t\right)>0$ lead to a non-trivial Wilson loop $\mathsf{ W}_1=\ee^{\ii2\pi\mathsf{ CS}_1}=-1$, signaling  a topological insulating ground-state. Note also that considering negative values of the   imbalance would lead to a similar topological  phase for $-4t<\Delta\epsilon<0$ with the role of the positive and negative Wilson masses interchanged $m_1>0, m_0<0$.
 
 \begin{figure}
\centering
\includegraphics[width=1.0\columnwidth]{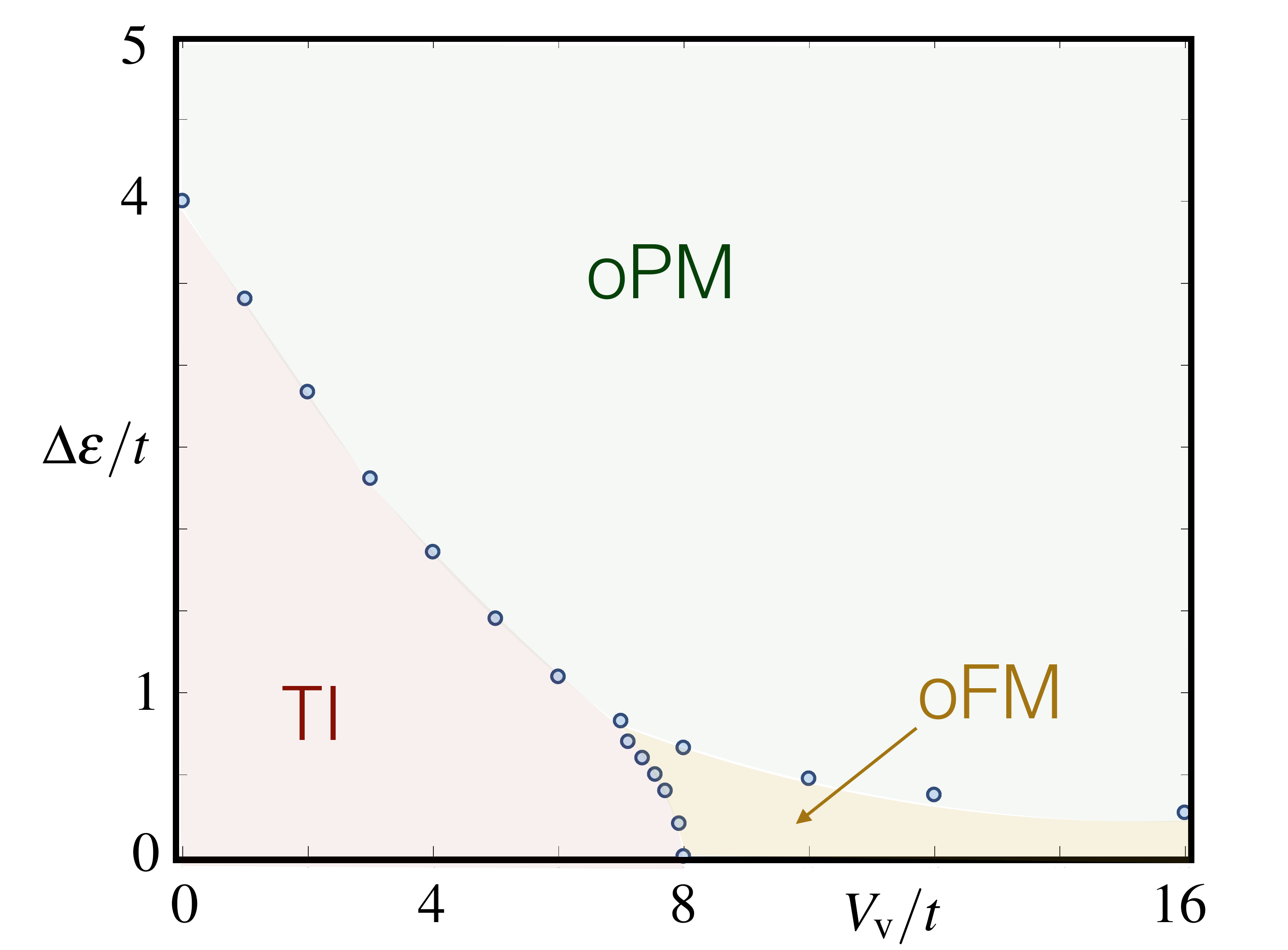}
\caption{ {\bf Phase diagram of the imbalanced Creutz-Hubbard ladder: } Phase diagram displaying a topological insulator (TI) phase, and two other non-topological phases, namely an orbital phase with long-range ferromagnetic Ising order (oFM), and an orbital paramagnetic phase (oPM).  The blue circles label numerical results, and the colored phase boundaries are a guide to the eye. }
\label{fig_phase_diagram}
\end{figure}

The presence of the energy imbalance breaks  both time-reversal $U_{\mathsf{T}} h_{\rm C}(-k)^* U_{\mathsf{T}}^\dagger = + h_{\rm C}(q)$ and particle-hole $U_{\mathsf{C}} h_{\rm C}(-k)^* U_{\mathsf{C}}^\dagger = - h_{\rm C}(k)$ symmetries,
 since the term $( \half\Delta \varepsilon + 2 {t}_{\rm x} \sin k)$ in Eq.~\eqref{eq:single_part_creutz} is neither even, nor odd,  under $k \leftrightarrow - k$. Hence,  the above symmetries cannot be fulfilled for any unitary $U_{\mathsf{T}}$ or  $U_{\mathsf{C}}$. On the other hand, there is a discrete sublattice symmetry  $U_{\mathsf{S}}h_{\rm C}({k})U_{\mathsf{S}}^\dagger=-h_{\rm C}({k})$ with the operator $U_{\mathsf{S}}=\sigma_y$.  Accordingly, the regime $\Delta\epsilon<4t$ with $\mathsf{ W}_1=-1$ can be interpreted as a 1D $\textsf{AIII}$ {\it  topological insulator}~\cite{creutz_hubbard,GNW_model_spt} .  
 
The goal of this section is to study the fate of this topological phase in the presence of 4-Fermi  terms
\beq
\label{eq:CH_ham}
H_{\rm CH}=H_{\rm C}+\frac{V_{\rm v}}{2}\sum_{j}\sum_\ell c_{j,\ell}^{\dagger}\,c_{j,\bar{\ell}}^{\dagger}\,c_{j,\bar{\ell}}^{\phantom{\dagger}}\,c_{j,\ell}^{\phantom{\dagger}},
\eeq
which can be understood as Hubbard density-density interactions~\eqref{eq:wilson_hubbard_ham} along the rungs of the ladder. In Ref.~\cite{creutz_hubbard}, we   obtained a numerical estimate of the phase diagram for this Creutz-Hubbard model using Matrix-Product-State methods~\cite{mps} (see Fig.~\ref{fig_phase_diagram}). In the present work, we will use these tools to benchmark the RG calculations, and test the validity of their connection to the topological Hamiltonian.
 
 \vspace{1ex}
{\it (b) Euclidean action and continuum  QFT description.--}
 \vspace{0.5ex}
 
 \noindent
In this subsection, we present the Euclidean action for the  continuum description of the Creutz-Hubbard ladder, which will be valid in the vicinity of the second-order quantum phase transition. We thus focus on  the vicinity of   $\Delta\epsilon= 4t$, where the Wilson fermion at $k_1=-\pi/2$ becomes massless, while the one at $k_0=+\pi/2$ has a large positive mass. We note that setting $t=t_{\rm x}$ makes this mass very heavy $m_0=2$, such that the corresponding Wilson fermion lies already at the  cutoff of the theory (i.e. maximum energy of the band). In any case, regardless of the particular value of $t/t_{\rm x}$, the fate of this fermion is to end in such a large mass limit as one approaches the IR limit of the RG transformations (i.e.  the mass term is a relevant perturbation~\eqref{eq:beta_mass}). We thus believe that, without loss of generality, one can set $t=t_{\rm x}$ from the outset.  Considering that this heavy fermion has a  big positive mass, the topological invariant~\eqref{eq:top_inv_CH} is fully controlled by the mass of the lighter fermion around $k_1$, such that a non-trivial integer-valued Wilson loop is obtained provided that
 \beq
 \label{eq:top_inv_CH_m0}
\mathsf{ W}_1=\ee^{\ii2\pi \mathsf{ CS}_1}=-1,\hspace{2ex}{\rm if}\hspace{1ex} {{\rm sgn}(m_{1})}=-1.
\eeq

We start by reorganizing the  action~\eqref{eq:Wilson_hubbard_action} for the Creutz-Hubbard Hamiltonian~\eqref{eq:CH_ham} in the regime $V_{\rm v}< \Delta\epsilon\approx 4t=4t_{\rm x}$  in a form that simplifies the RG calculations beyond tree level~\eqref{eq:beta_mass}-\eqref{eq:hubb_tree_level}, namely $S_{\rm CH}=S_0+\delta S$ with 
\beq
\label{eq:S0_ch}
S_0=\bigintssss_{\{{k},\omega\}}\sum_{\eta={\rm R,L,u,d}}\overline{\psi}_{\eta}({k},\omega)(-\ii\omega+\epsilon_{\eta}({k}))\psi_{\eta}^{\phantom{\dagger}}({k},\omega),
\eeq
where we have introduced the flavor index $\eta$ to label the different Wilson fermions: {\it (i)} $\eta\in\{{\rm R,L}\}$ refers to the right- ($s_{\rm R}=1$) as  and left-moving ($s_{\rm L}=-1$) modes around the Dirac point $k_1$, $\psi_{\eta}^{\phantom{\dagger}}({k},\omega)=(\psi_{1,\rm u}^{\phantom{\dagger}}({k},\omega)-s_{\eta}\psi_{1,\rm d}^{\phantom{\dagger}}({k},\omega)/\sqrt{2}$, which have energies  $\epsilon_{\eta}({k})=s_\eta k$. The  Wilson-mass term of the fermions  around this point $m_1$, which is small for $\Delta\epsilon\approx 4t$, will be included in $\delta S$. In addition, {\it(ii)} we define the flavors  $\eta\in\{{\rm u,d}\}$ for the positive- and negative-frequency modes around the point $k_0$,  $\psi_{\eta}^{\phantom{\dagger}}({k},\omega)=\psi_{0,\eta}^{\phantom{\dagger}}({k},\omega)$, which have  energies $\epsilon_{\eta}({k})=\pm m_{0}$. We note that the corrections to this heavy-mass limit (i.e. $\delta\epsilon_{\eta}({k})\approx \mp  k^2$)
 can be included in $\delta S$. However,  as occurred for the shorter-wavelength corrections~\eqref{eq:short_wavelength_action}, these perturbations  are IR irrelevant   already at tree level.

  \begin{figure}
\centering
\includegraphics[width=1.0\columnwidth]{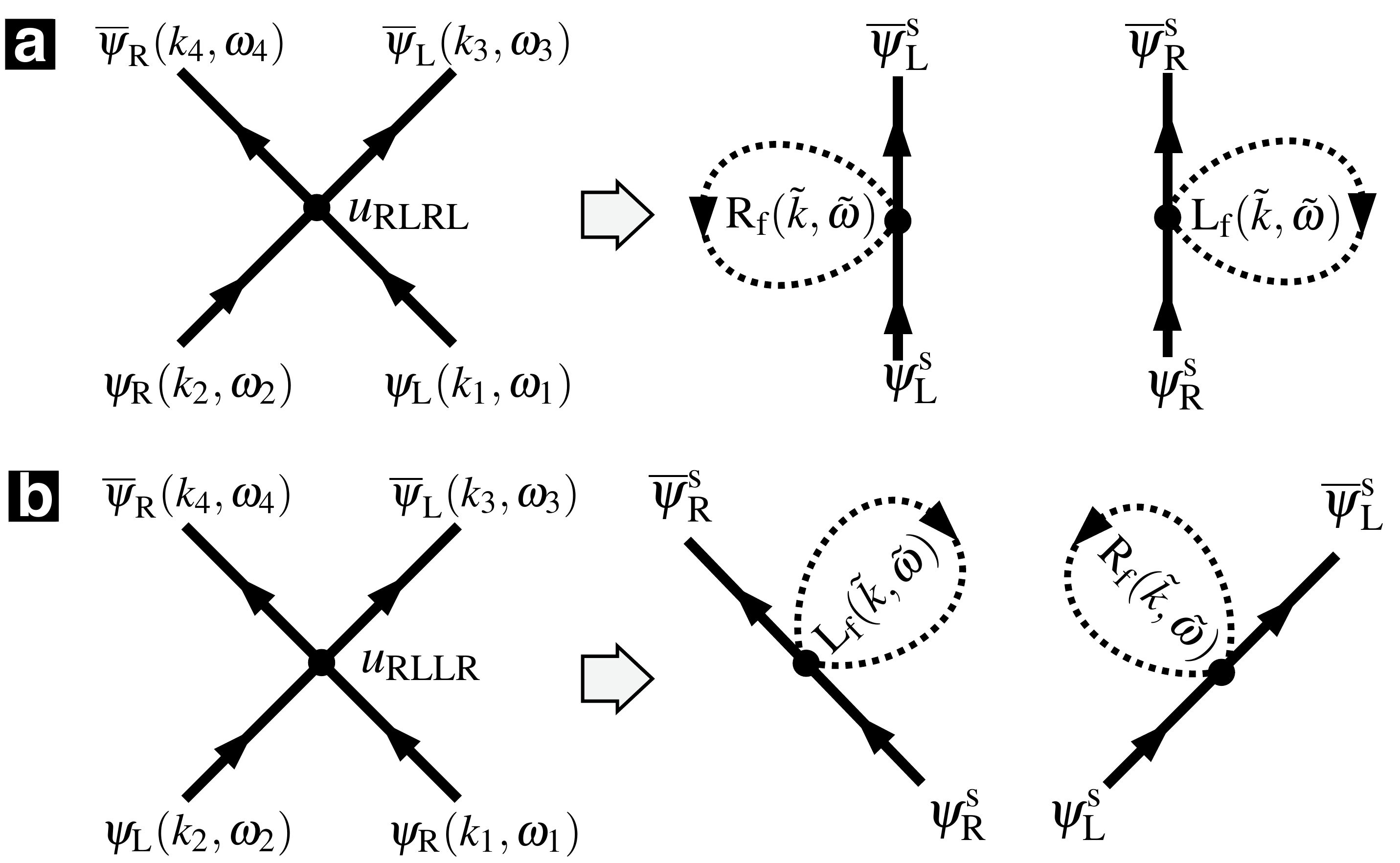}
\caption{ {\bf One-loop tadpole diagrams around  $k_1$: } {\bf (a)} (left) Forward scattering  of $(\eta_1,\eta_2)=({\rm L,R})$ fermions onto $(\eta_3,\eta_4)=({\rm L,R})$. (right) Tadpole diagrams by joining an incoming and outgoing line of the same $\eta\in\{\rm L,R\}$ flavor into a loop. The momentum inside the loop is constrained to a small region around the cutoff $\Lambda_{\rm c}/s\leq \tilde{k}\leq\Lambda_{\rm c}$. The loop integrals  over the fast modes   are labelled by $\eta_{\rm f}(\tilde{k},\tilde{\omega})$ and depicted by  dashed closed  lines. {\bf (b)} (left) Backward scattering  of $(\eta_1,\eta_2)=({\rm R,L})$ fermions onto $(\eta_3,\eta_4)=({\rm L,R})$. (right)  Corresponding tadpole diagrams.}
\label{fig_tadpoles_K1}
\end{figure}

 The  perturbations that are relevant and marginal at tree level are contained in $\delta S=\delta S_{\rm m}+\delta S_{\rm int}$, where 
 \beq
 \label{eq:Sm_ch}
 \delta S_{\rm m}=\bigintssss_{\{{k},\omega\}}\sum_{\eta={\rm L, R}}m_1\Lambda_{\rm c}\overline{\psi}_{\eta}({k},\omega)\psi_{\bar{\eta}}^{\phantom{\dagger}}({k},\omega),
 \eeq
 and we have introduced $\bar{\eta}={\rm L}$ ($\bar{\eta}={\rm R}$) for ${\eta}={\rm R}$ (${\eta}={\rm L}$). Clearly, the mass term mixes the right- and left-moving fermions around $k_1$, breaking explicitly the chiral symmetry  of the free masless Dirac fermion~\eqref{eq:S0_ch}~\cite{fradkin_book}. The correction due to the Hubbard interaction~\eqref{eq:CH_ham}, which must be normalized by the effective speed of light $c=2t_{\rm x}=2t$, can be expressed in a form similar  to Eq.~\eqref{eq:S_int}, namely
\beq
\label{eq:int_action_CH}
\delta S_{\rm int}=-\frac{1}{4}\bigintsss_{\{{k},\omega\}}\!\sum_{\{\boldsymbol{\eta}\}}u_{\boldsymbol{\eta}}\overline{\psi}_{\eta_4}(4)\overline{\psi}_{\eta_3}(3)\psi_{\eta_2}(2)\psi_{\eta_1}(1),
\eeq
where we use  similar conventions as below Eq.~\eqref{eq:S_int}, and define $\sum_{\{\boldsymbol{\eta}\}}=\sum_{\eta_1,\eta_2,\eta_3,\eta_4}$. These interaction terms describe the scattering between an  incoming pair of fermions $(\eta_1,\eta_2)$  onto an outgoing pair $(\eta_3,\eta_4)$. 
Note that the dimensionless couplings $u_{\boldsymbol{\eta}}=u_{\eta_4,\eta_3,\eta_2,\eta_1}$ should be antisymmetric with respect to $\eta_4\leftrightarrow\eta_3$, or  $\eta_2\leftrightarrow\eta_1$, as a consequence of the anti-commuting nature of the Grassmann variables~\cite{shankar_rg}. Additionally, the particular form of the Hubbard interaction in Eq.~\eqref{eq:CH_ham} leads to couplings $u_{\boldsymbol{\eta}}$ without any momentum dependence. 

In combination, these two features limit the possible scattering events   underlying the action~\eqref{eq:int_action_CH}. For instance, for scattering processes that take place in the vicinity of a single  point $k_n$, the only allowed couplings are $u_{\rm RLLR}:=V_{\rm v}/2t$, and $u_{\rm uddu}:=2V_{\rm v}/t$, together with all the possible permutations

\beq
\begin{split}
\label{eq:couplings_1}
u_{\rm RLLR}&=-u_{\rm LRLR}=-u_{\rm RLRL}=u_{\rm LRRL}:=V_{\rm v}/2t,\\
u_{\rm uddu}&=-u_{\rm udud}\,\,=\,\,-u_{\rm dudu}\,\,=\,\,u_{\rm duud}:=2V_{\rm v}/t,\\
\end{split}
\eeq
 All these scattering processes preserve the number of fermions with a given  flavor $\eta$ (see the left panels of Figs.~\ref{fig_tadpoles_K1} {\bf(a)} and {\bf(b)} for the two possible scattering channels with an outgoing $(\eta_3,\eta_4)=(\rm L,R)$ pair).
 
\begin{figure}
\centering
\includegraphics[width=1.0\columnwidth]{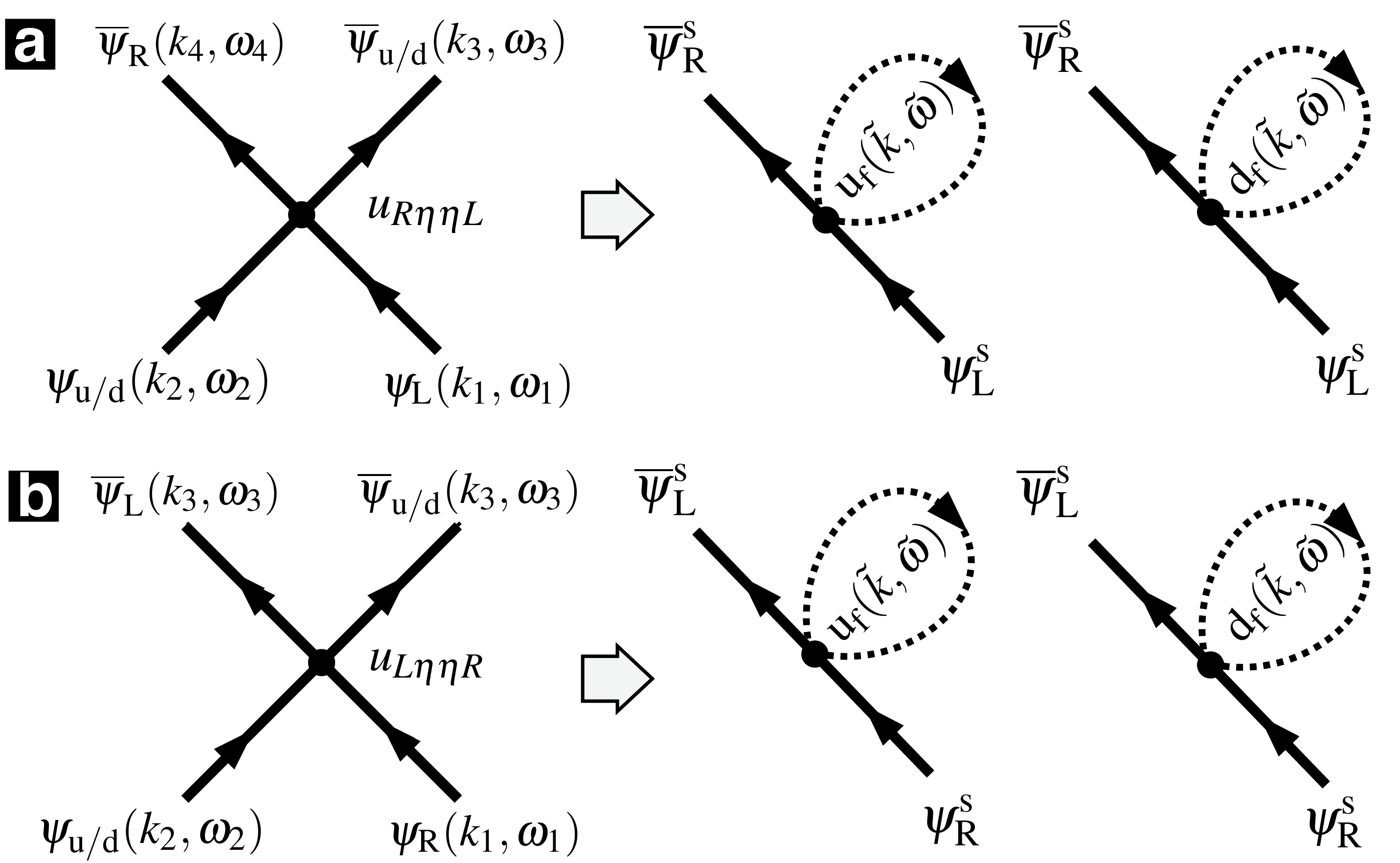}
\caption{ {\bf One-loop tadpole diagrams involving  $k_1$ and $k_0$: } {\bf (a)} (left)  Scattering  of $(\eta_1,\eta_2)=({\rm L,u/d})$ fermions onto $(\eta_3,\eta_4)=({\rm u/d,R})$. (right) Tadpole corrections to the mass of the light fermions  by joining an incoming and outgoing line of the heavy fermions (we use the same conventions as in Fig.~\ref{fig_tadpoles_K1}) . {\bf (b)} Same as for {\bf (a)} but for the scattering  of $(\eta_1,\eta_2)=({\rm R,u/d})$ fermions onto $(\eta_3,\eta_4)=({\rm u/d,L})$.}
\label{fig_tadpoles_K1K0}
\end{figure}

In addition, there will be scattering processes that involve fermions from both  Dirac points  $k_0$ and  $k_1$, which can be organized in two  sets. The first set consists of   scattering processes that conserve the number of fermions around each point, but may change their flavor e.g. a $L\leftrightarrow R$ and $u\leftrightarrow d$, namely 
\beq
\begin{split}
\label{eq:couplings_2}
u_{\rm RuuR}&=-u_{\rm RuuL}=-u_{\rm LuuR}=u_{\rm LuuL}:=V_{\rm v}/4t,\\
u_{\rm RddR}&=\phantom{-}u_{\rm RddL}=\phantom{-}u_{\rm LddR}=u_{\rm LddL}:=V_{\rm v}/4t,\\
u_{\rm LudL}&=-u_{\rm  RudR}=-u_{\rm RudL}=u_{\rm LudR }:=V_{\rm v}/4t,\\
u_{\rm  LduL}&=-u_{\rm RduR}=-u_{\rm LduR }=u_{\rm  RduL}:=V_{\rm v}/4t,\\
\end{split}
\eeq
plus all permutations $\eta_1\eta_2$, $\eta_3\eta_4$ with their relative signs. 
The second set corresponds to the so-called Umklapp scattering, i.e. processes where the number of fermions around each point  changes in pairs
\beq
\begin{split}
\label{eq:couplings_3}
u_{\rm udRL}&=-u_{\rm udLR}=-u_{\rm duRL}=u_{\rm duLR}:=V_{\rm v}/t,\\
u_{\rm LRdu}&=-u_{\rm LRud}=-u_{\rm RLud}=u_{\rm RLud}:=V_{\rm v}/t.\\
\end{split}
\eeq
We note that momentum conservation can still be achieved  up to a reciprocal lattice vector  $\pm 2(k_1-k_0)=\pm2\pi$.

Equations~\eqref{eq:S0_ch}-\eqref{eq:int_action_CH} set the stage for the RG study of correlation effects in the  one-dimensional  $\textsf{AIII}$ topological insulator beyond tree level. As discussed in detail below, in the vicinity of   $\Delta\epsilon= 4t=4t_{\rm x}$, we shall use these corrections to predict the critical line separating topological and normal  insulating phases at finite repulsive interactions  $V_{\rm v}>0$.

\subsection{Loop-expansion:  running of the Wilson masses and  topological invariants in presence of interactions}
\label{subsec:loop_expansion_ch}

In this subsection, we consider the first- and second-order terms of the cumulant expansion~\eqref{eq:cum_exp_action} of the quartic interactions~\eqref{eq:int_action_CH}, focussing on its effects on the renormalization of the Wilson masses and the topological invariant. This will allow us to determine the critical line that connects to the non-interacting RG fixed point  separating the topological and trivial insulators, and to show that such a critical line delimits the region of the phase diagram where the ground-state corresponds to  a correlated topological insulator.  

 \vspace{1ex}
{\it (a) Vanishing  tadpoles considering  light Wilson fermions.--}
 \vspace{0.5ex}
 
 \noindent
In the regime $\Delta\epsilon\approx 4t=4t_{\rm x}$,  the Wilson masses fulfil $|m_1|\ll |m_0|$, and one may expect that the heavy fermions around $k_0$  shall not have any influence on the lighter fermions around $k_1$, which in turn  determine  the onset of a topological phase (i.e. the mass-inversion point  $m_1<0$). Following this line of reasoning,  to see how this Wilson mass runs with the cutoff $m_1(\Lambda_{\rm c})$, and determine how the RG fixed point ${m_1}=0$ changes with interactions, it would suffice to consider scattering events~\eqref{eq:int_action_CH} in the vicinity of $k_1$, i.e. scattering between left- and right-moving modes described by  the first line of Eq.~\eqref{eq:couplings_1}.  The first-cumulant correction to the coarse-grained action would then arise from the following terms 
\begin{widetext} 
\beq
\label{single_K_correction_one_loop}
\begin{split}
\langle \delta  S^{k_1}_{\rm int}\rangle _{\rm f}=&\frac{1}{4}\bigintsss_{\{{k},\omega\}}\!\sum_{{\eta}={\rm L,R}}u_{\eta\bar{\eta}\bar{\eta}\eta}\left(\left\langle\overline{\psi}_{\bar{\eta}}^{\rm f}(3)\psi_{\bar{\eta}}^{\rm f}(2)\right\rangle_{\rm f}\overline{\psi}_{\eta}^{\rm s}(4)\psi_{\eta}^{\rm s}(1)+\left\langle\overline{\psi}_{{\eta}}^{\rm f}(4)\psi_{{\eta}}^{\rm f}(1)\right\rangle_{\rm f}\overline{\psi}_{\bar{\eta}}^{\rm s}(3)\psi_{\bar{\eta}}^{\rm s}(2)\right)\\
+&\frac{1}{4}\bigintsss_{\{{k},\omega\}}\!\sum_{{\eta}={\rm L,R}}u_{\eta\bar{\eta}{\eta}\bar{\eta}}\left(\left\langle\overline{\psi}_{\bar{\eta}}^{\rm f}(3)\psi_{\bar{\eta}}^{\rm f}(1)\right\rangle_{\rm f}\overline{\psi}_{\eta}^{\rm s}(4)\psi_{\eta}^{\rm s}(2)+\left\langle\overline{\psi}_{{\eta}}^{\rm f}(4)\psi_{{\eta}}^{\rm f}(2)\right\rangle_{\rm f}\overline{\psi}_{\bar{\eta}}^{\rm s}(3)\psi_{\bar{\eta}}^{\rm s}(1)\right),\\
\end{split}
\eeq
\end{widetext} 
where $\langle\overline{\psi}_{{\eta}}^{\rm f}(j)\psi_{{\eta}}^{\rm f}(j')\rangle_{\rm f}=\delta(\omega_j-\omega_{j'})\delta(k-{k'})/(\ii\omega_j- \epsilon_\eta (k))$ is obtained from the free-fermion propagator~\eqref{eq:green_function_free} by working on the corresponding eigenbasis of the non-interacting single-particle Hamiltonian.

These contributions  can be depicted in terms of the so-called {\it tadpole diagrams} (see the right panels of Figs.~\ref{fig_tadpoles_K1} {\bf(a)} and {\bf(b)} for $\eta={\rm L}$), which include a single closed loop over the fast modes. It is already apparent from the fast-mode loops of these figures, without any further calculation, that these terms  can only contribute with $\langle \delta  S^{k_1}_{\rm int}\rangle _{\rm f}\propto (\overline{\psi}_{\rm R}^{\rm s}(k,\omega)\psi_{\rm R}^{\rm s}(k,\omega)+\overline{\psi}_{\rm L}^{\rm s}(k,\omega)\psi_{\rm L}^{\rm s}(k,\omega))$, which is an irrelevant common shift of the on-sites energies for the slow modes. However, there is no  one-loop term that yields a correction to the  mass term  mixing the different fermion chiralities $\langle \delta  S^{k_1}_{\rm int}\rangle _{\rm f}\propto( \overline{\psi}_{\rm R}^{\rm s}(k,\omega)\psi_{\rm L}^{\rm s}(k,\omega)+\overline{\psi}_{\rm L}^{\rm s}(k,\omega)\psi_{\rm R}^{\rm s}(k,\omega))$, inducing thus a non-vanishing  mass~\eqref{eq:Sm_ch} around $k_1$. 
This result is in contradiction to the numerical data displayed in Fig.~\ref{fig_phase_diagram}, where the non-interacting critical point  at $\Delta\epsilon= 4t$ flows towards smaller values of the imbalance linearly  $(\Delta\epsilon- 4t)\propto -V_{\rm v}$, for $V_{\rm v}\ll\Delta\epsilon$. Accordingly, the Wilson mass  corresponding to this fixed point $m_1$ should get a renormalization $\tilde{m}_1=m_1+\delta m_1(V_{\rm v})$~\eqref{dressed_masses}  linear in the interaction strength $\delta m_1(V_{\rm v})\propto V_{\rm v}$. We can thus conclude that, when restricting to the slow modes of the continuum description, it is not possible to obtain non-zero tadpole corrections that can account for the numerical phase diagram.

We note, at this point, that 4-Fermi continuum QFTs can display the phenomenon of dynamical mass generation, where an interacting massless Dirac fermion  acquires a mass that spontaneously breaks chiral symmetry~\cite{GN_model}. The scaling of the dynamically-generated mass with the interaction strength is, however, non-perturbative~\cite{GN_model, GNW_model_spt}, and cannot account for the aforementioned linear dependence. We will extend on the interplay of lattice effects and  a dynamically-generated mass in the last section with our conclusions and outlook. 

 \vspace{1ex}
{\it (b) Tadpoles considering also heavy Wilson fermions.--}
 \vspace{0.5ex}
 
Let us now explore how a linear running of the Wilson mass can be obtained by considering the effect that  the  heavy fermion around $k_0$ may have  on the lighter  mass $m_1$ of  the fermion around  $k_1$. The heavy fermion around $k_0$, lying at the cutoff of the theory, may indeed renormalize the parameters of the QFT when integrated out during the RG coarse-graining.
Inspecting Eqs.~\eqref{eq:couplings_2}-\eqref{eq:couplings_3}, one  realizes that  the couplings $u_{\rm LuuR},u_{\rm RuuL},u_{\rm LddR},u_{\rm RddL}$, together with their corresponding permutations, can lead to tadpole diagrams that indeed induce  a correction to the mass (see Fig.~\ref{fig_tadpoles_K1K0}). The  first-cumulant contribution to the coarse-grained  action  yields
\begin{widetext} 
\beq
\label{eq:mass_term_action}
\begin{split}
\langle \delta  S^{k_1k_0}_{\rm int}\rangle _{\rm f}=&\frac{1}{4}\bigintsss_{\{{k},\omega\}}\!\sum_{{\eta}={\rm u,d}}4\left(u_{\rm R{\eta}{\eta}L}\left\langle\overline{\psi}_{\eta}^{\rm f}(4)\psi_{{\eta}}^{\rm f}(1)\right\rangle_{\!\!\rm f}\overline{\psi}_{\rm R}^{\rm s}(3)\psi_{\rm L}^{\rm s}(2)+u_{\rm L{\eta}{\eta}R}\left\langle\overline{\psi}_{\eta}^{\rm f}(4)\psi_{{\eta}}^{\rm f}(1)\right\rangle_{\!\!\rm f}\overline{\psi}_{\rm L}^{\rm s}(3)\psi_{\rm R}^{\rm s}(2)\right),\\
\end{split}
\eeq
\end{widetext}
where the additional factor of $4$ comes from considering the contribution of all  possible permutations, e.g. $u_{\rm LuuR}=-u_{\rm uLuR}=-u_{\rm LuRu}=u_{\rm uLRu}$, all of which contribute equally. 

\begin{figure}
\centering
\includegraphics[width=1\columnwidth]{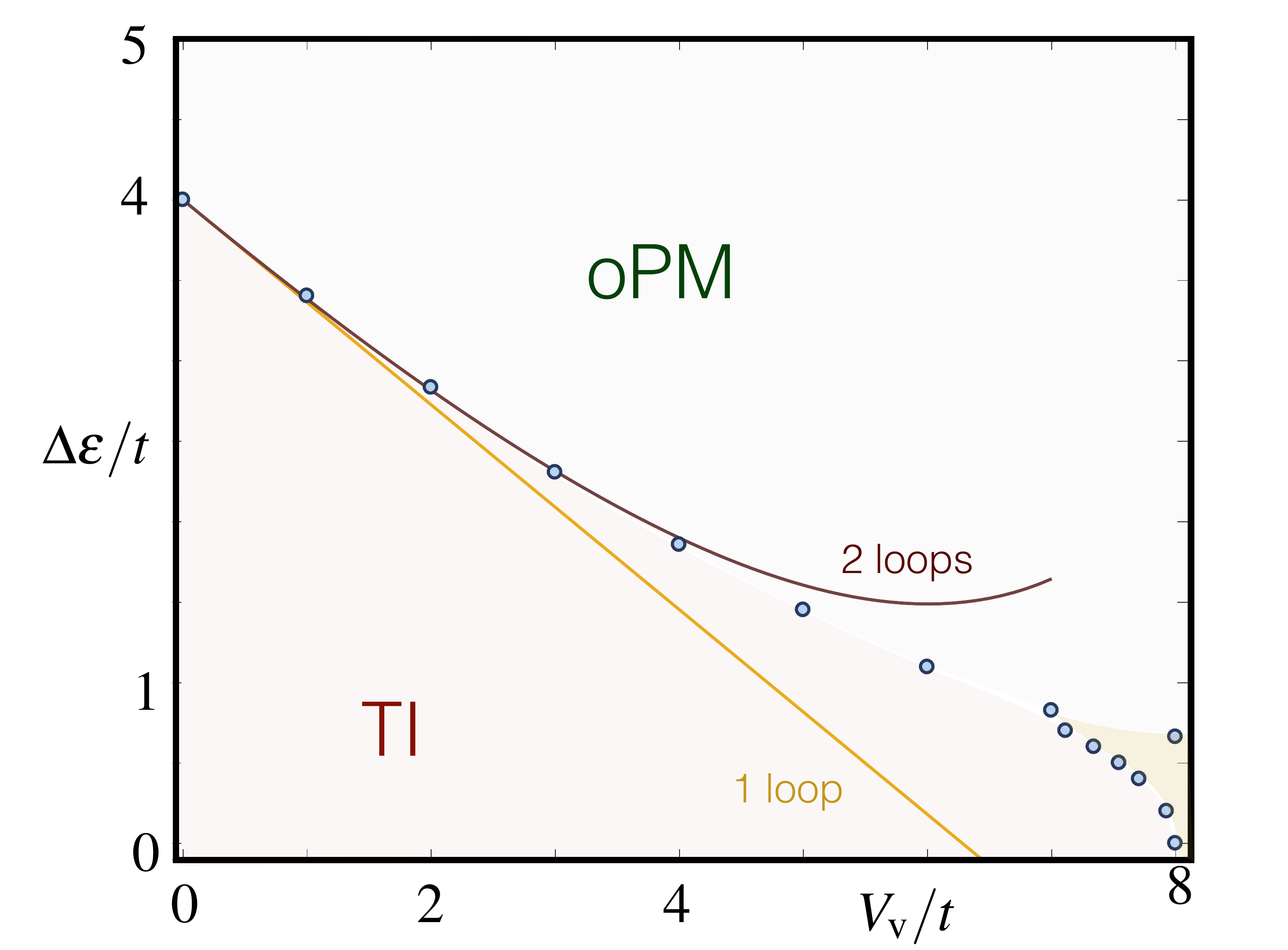}
\caption{ {\bf Benchmark of the RG prediction for the critical line: } Comparison of the one-loop~\eqref{eq:critical_point} (yellow line) and two-loop~\eqref{eq:two_loop_critical_line} (red line) predictions of the critical line. The blue circles label numerical results of Fig.~\ref{fig_phase_diagram} in the region of  weak to moderate interactions. }
\label{fig_phase_diagram_zoom}
\end{figure}

If we now perform the loop integrals, we obtain 
$
\int_{\mathbb{R}}\!\!\frac{{\rm d}\omega_4}{2\pi}\!\int_{\frac{\Lambda_{\rm c}}{s}}^{\Lambda_{\rm c}}\!\!\frac{{\rm d}k_4}{(2\pi)}\int_{\mathbb{R}}\!\!\frac{{\rm d}\omega_1}{2\pi}\int_{\frac{{\Lambda_{\rm c}}}{s}}^{\Lambda_{\rm c}}\!\!\frac{{\rm d}k_1}{(2\pi)}\left\langle\overline{\psi}_{\eta}^{\rm f}(4)\psi_{{\eta}}^{\rm f}(1)\right\rangle_{\rm f}=\mp\frac{\Lambda_{\rm c}}{\pi}(1-\frac{1}{s})
$ for $\eta={\rm u,d}$, respectively. To compare with the original action, we must now rescale the momentum and frequency~\eqref{eq:rescaling} to reset the original cutoff. Accordingly, the slow modes should also be transformed  
\beq
\begin{split}
\psi^{\rm s}_{\rm R/L}({k},\omega)\to{\psi}'_{\rm R/L}({k}',\omega')=s^{-(d+2)/2}{\psi}^{\rm s}_{\rm R/L}({k}'/s,\omega'/s),
\end{split}
\eeq
such that the first-order cumulant~\eqref{eq:mass_term_action} contributes with a correction to the Wilson mass~\eqref{eq:Sm_ch}, namely 
 \beq
\langle \delta  S^{k_1k_0}_{\rm int}\rangle _{\rm f}=\bigintssss_{{k},\omega}\sum_{\eta={\rm L, R}}\delta m_1^{(1)}\Lambda_{{\rm c}}\overline{\psi}_{\eta}({k},\omega)\psi_{\bar{\eta}}^{\phantom{\dagger}}({k},\omega).
 \eeq
 Here, we have introduced the  Wilson mass shift  to one loop $ \delta m_1^{(1)}$, which modifies the  $\beta_{m_1}$ function in the following way
 \beq
 \label{eq:ruuning_mass}
 \delta m_1^{(1)}=\frac{V_{\rm v}}{2\pi t}(s-1),\hspace{3ex} \beta_{m_1}=\frac{{\rm d}m_{1}}{{\rm d}\log s}=m_1+ \delta \beta_{m_1}^{(1)}.
 \eeq
 
 The RG fixed point is then determined by the bare lattice parameters that lead to a vanishing $\beta_{m_1}$ function, namely
 \beq
 \label{eq:critical_point}
 \frac{\Delta\epsilon}{4t}-1=-\frac{V_{\rm v}}{2\pi t}\implies\left.\frac{\Delta\epsilon}{t}\right|_{\rm c}=4-\frac{2}{\pi}\frac{V_{\rm v}}{t}.
 \eeq
The  comparison of this prediction with the numerical results shows a good agreement in the regime of weak interactions (see the yellow line in Fig.~\ref{fig_phase_diagram_zoom}). We note that this one-loop correction to the mass agrees exactly with a self-consistent mean-field treatment that relies on the mapping of the Creutz-Hubbard model to a pair of coupled quantum Ising models~\cite{creutz_hubbard}. The advantage of the present RG approach is that, on the hand, it can be improved systematically by considering higher orders in the cumulant expansion. On the other hand,  as discussed below, we can not only predict the position of the critical line, but also show that the region it delimits corresponds to a correlated $\mathsf{AIII}$ topological insulator.
 
\vspace{1ex}
{\it (c) Two-loop corrections to the light Wilson mass.--}
\vspace{0.5ex}

\noindent
Let us now move on to the second-order cumulant contributions~\eqref{eq:cum_exp_action} to the coarse-grained action, which will yield two-loop corrections to the Wilson mass that can
be accounted for by considering the so-called one-particle irreducible Feynman diagrams with two interaction vertices and two external lines with right- and left-moving fermions.  

\begin{figure}
\centering
\includegraphics[width=1.0\columnwidth]{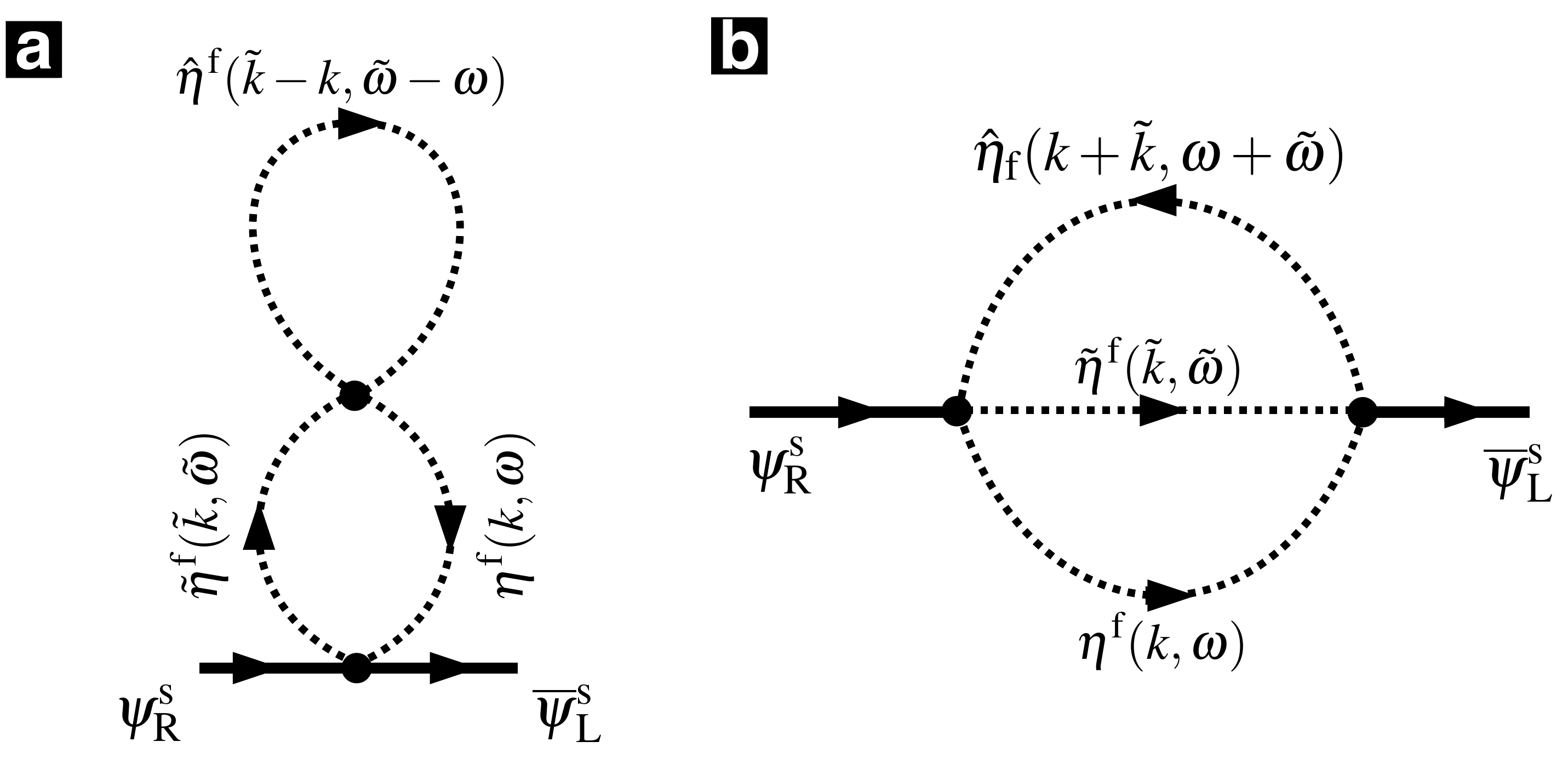}
\caption{ {\bf  Two-loop diagrams for the Wilson mass term: } {\bf (a)-(b)} Possible two-loop diagrams for the correction to the mass of the light fermions $\eta_j\in\{\rm L,R\}$ by virtual creation-annihilation of light/heavy fermions in fast modes $\eta,\tilde{\eta},\hat{\eta}\in\{\rm R,L,u,d\}$.}
\label{fig_mass_diagrams_2_loops}
\end{figure}

In analogy to the discussion below Eq.~\eqref{single_K_correction_one_loop}, we note that it is not possible to obtain two-loop corrections to the mass by simply  focusing on the light fermion around $k_1$ (i.e.   scattering events in first line of Eq.~\eqref{eq:couplings_1}). By incorporating  the interactions with the heavy fermion around $k_0$, we can have the contributions to the Wilson mass depicted in  Fig.~\ref{fig_mass_diagrams_2_loops}, where we recall that the closed loops are formed by heavy-fermion propagators. In contrast to the standard situation in other interacting QFTs~\cite{amit_book}, where the double tadpole of Fig.~\ref{fig_mass_diagrams_2_loops} {\bf (a)} would  contribute to the mass, here  
we  find that only the ``saturn diagram'' of Fig.~\ref{fig_mass_diagrams_2_loops} {\bf (b)} does  contribute with a second-order shift of the mass
 \beq
\left\langle   \left(\delta S^{K_1K_0}_{\rm int}\right)^2\right\rangle _{\!\!\rm c, f}\!=\bigintssss_{{k},\omega}\sum_{\eta={\rm L, R}}\delta {m}^{(2)}_1\Lambda_{{\rm c}}\overline{\psi}_{\eta}({k},\omega)\psi_{\bar{\eta}}^{\phantom{\dagger}}({k},\omega).
 \eeq
 Here, we have introduced the two-loop mass shift
 \beq
 \delta{m}_1^{(2)}=-\frac{3}{8\pi^2}\frac{V_{\rm v}^2}{m_0t^2}(s-1), \hspace{3ex} \beta_{m_1}=m_1+ \delta \beta_{m_1}^{(1)}+\delta\beta _{{m}_1}^{(2)}.
 \eeq
 which should be evaluated for the mass of the heavy fermions  $m_0=\Delta\epsilon/4t+1$  at the bare lattice parameters that yield the RG fixed point with first-order corrections~\eqref{eq:critical_point}, such that  $m_0=2-V_{\rm v}/2\pi t$. Hence, the critical line of the model with two-loop corrections follows from the condition of a vanishing $\beta_{m_1}$ function, and yields
 \beq
 \label{eq:two_loop_critical_line}
 \left.\frac{\Delta\epsilon}{t}\right|_{\rm c}=4-2\left(\frac{V_{\rm v}}{\pi t}\right)+\frac{3}{4}\left(\frac{V_{\rm v}}{\pi t}\right)^2\frac{1}{4-\left(\frac{V_{\rm v}}{\pi t}\right)}.
 \eeq
 The  comparison of this prediction with the numerical results shows a much better  agreement in the regime of weak to intermediate interactions (see the red line of Fig.~\ref{fig_phase_diagram_zoom}).
 
 \vspace{1ex}
{\it (d) RG flows and the effective topological Hamiltonian.--}
 \vspace{0.5ex}

\noindent
 Let us now consider the nature of the two phases separated by this critical line from the perspective of the topological Hamiltonian~\eqref{top_ham}. As discussed below Eq.~\eqref{top_ham}, the renormalization of the Wilson masses~\eqref{dressed_masses} contributes to the zero-frequency self-energy in a very simple manner, which allows to diagonalize the topological Hamiltonian~\eqref{top_ham} in complete analogy to the single-particle one. In the present case, the topological Hamiltonian  can be expressed as
\beq
\label{eq:single_part_creutz2}
h_{\rm top}({k})=-2t\sigma^x\cos k+\left(\half\Delta\tilde{\epsilon}+2t\sin k\right)\sigma^z,
\eeq
where the  previous two-loop corrections contribute to
\beq
\label{eq:RG_imbalance}
\Delta\tilde{\epsilon}=\Delta\epsilon+\frac{2}{\pi}V_{\rm v} -\frac{3V_{\rm v}}{4\pi^2 t}\left(4-\frac{V_{\rm v}}{\pi t}\right).
\eeq

The calculation  of the topological invariants now directly yields Eq.~\eqref{eq:CS_AIII_interacting}, and we obtain a non-zero topological invariant for the correlated topological insulator when 
\beq
\label{eq:top_inv_CH_m0_int}
\mathsf{ W}_1=\ee^{\ii2\pi \mathsf{ CS}_1}=-1,\hspace{2ex}{\rm if}\hspace{1ex} {{\rm sgn}(\tilde{m}_{1})}=-1.
\eeq
Since we know the two-loop corrections to the mass, we can  characterize the nature of the insulating phases separated by the critical line:  {\it (i)} for $\frac{\Delta\epsilon}{t}> \left.\frac{\Delta\epsilon}{t}\right|_{\rm c}$ (i.e. above the critical line), the mass  $\tilde{m}_1>0$, such that $\mathsf{CS}_1=0$, and the Wilson loop is trivial $\mathsf{W}_1=1$. Accordingly, the ground-state corresponds to a normal band insulator.  {\it (ii)} for $\frac{\Delta\epsilon}{t}< \left.\frac{\Delta\epsilon}{t}\right|_{\rm c}$ (i.e. below the critical line), the mass  $\tilde{m}_1<0$, such that $\mathsf{CS}_1=\half$, and the Wilson loop is  $\mathsf{W}_1=-1$. Accordingly, the ground-state corresponds to a topological $\mathsf{AIII}$ insulator. This behaviour is in complete agreement with  Fig.~\ref{fig_phase_diagram}, and thus can serve as a quantitative test of the validity of the RG picture of correlated topological insulators presented in this work.

\begin{figure}
\centering
\includegraphics[width=0.9\columnwidth]{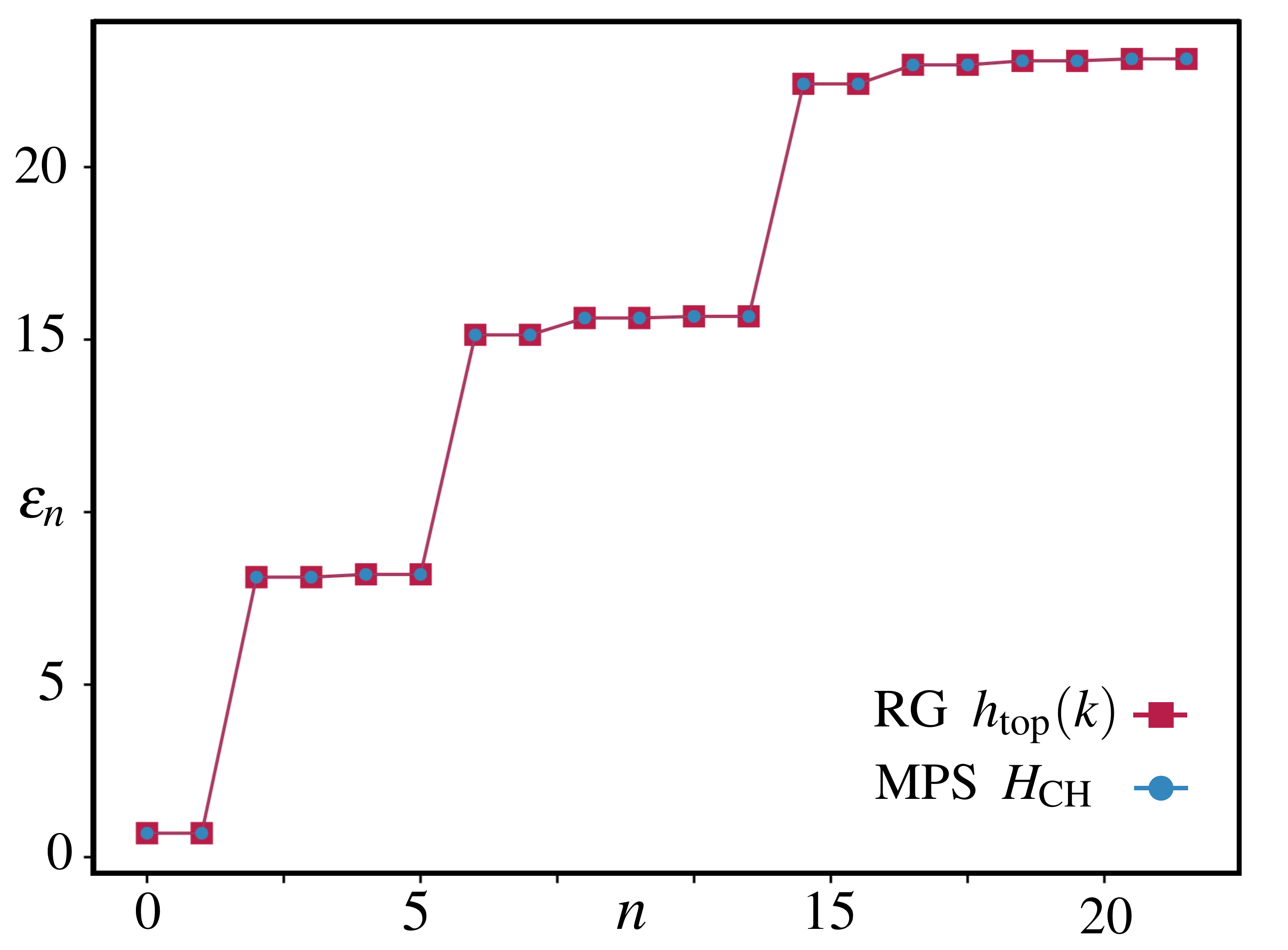}
\caption{ {\bf  Entanglement spectrum of the Creutz-Hubbard ladder: } Lowest $N_{\rm ES}=22$ eigenvalues of the entanglement spectrum $\epsilon_n\in {\rm ES}(\rho_\ell)$ for the Creutz-Hubbard ladder~\eqref{eq:CH_ham} for $\Delta\epsilon=2.2t$, $V_{\rm v}=0.2 t$, which lies inside the TI region of Fig.~\ref{fig_phase_diagram_zoom}. We compare the results obtained via the MPS numerical simulations (squares), and via the RG-corrected topological Hamiltonian (circles) ~\eqref{eq:single_part_creutz2}.}
\label{fig_ES_comparison}
\end{figure}

\begin{figure*}
\centering
\includegraphics[width=2.\columnwidth]{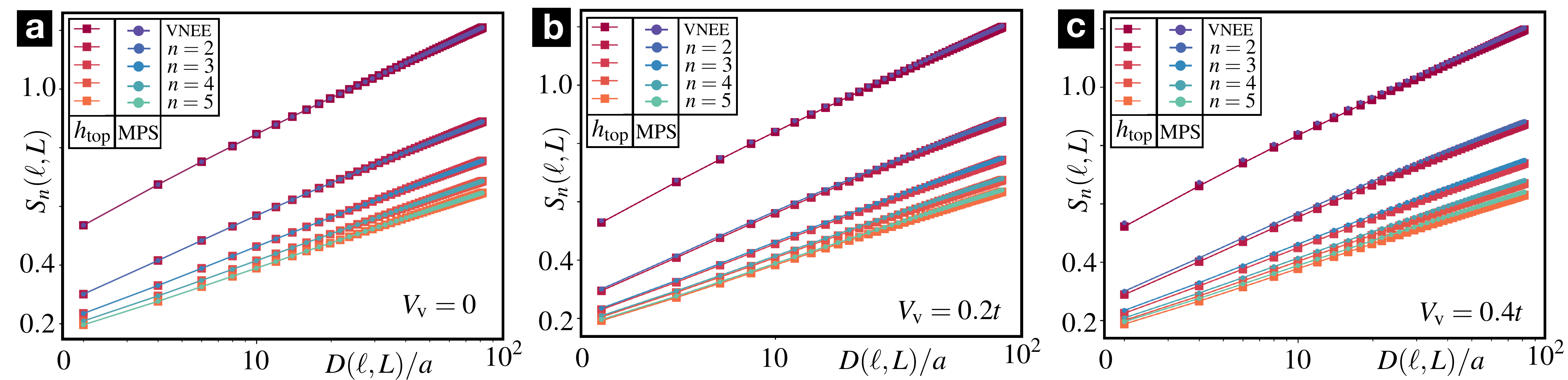}
\caption{ {\bf  Entanglement entropies of the critical Creutz-Hubbard ladder: } Von Neumann and R\'enyi block entanglement entropies are represented as a function of the block chord length. The Hubbard interaction strengths are set to {\bf (a)} $V_{\rm v}=0$, {\bf (b)} $V_{\rm v}=0.2t$, and {\bf (c)} $V_{\rm v}=0.4t$, while the corresponding energy imbalance $\Delta\epsilon$  is set to the critical value of Eq.~\eqref{eq:two_loop_critical_line} predicted by the RG calculations.}
\label{fig_REEs_comparison}
\end{figure*}

\subsection{Entanglement spectroscopy: symmetry-protected topological phases and critical Luttinger liquids}
\label{subsec:entanglement_spect}

In this subsection, we shall explore additional features of the zero-temperature Creutz-Hubbard  ladder to benchmark the     RG-corrected topological Hamiltonian~\eqref{eq:single_part_creutz}. These features will become manifest in  the  bipartite correlations of a ground-state partitioned into two blocks of length $\ell$ and $L-\ell$, where $L=Na$ is the length of the whole ladder. In particular, we shall be interested in two types of correlations: entanglement entropies, and bi-partite fluctuations. The former are related to the so-called entanglement spectrum, which serves to characterize the topological properties of an interacting topological insulator as a symmetry-protected topological (SPT) phase. The latter will be used to characterize the critical line delimiting the topological-insulating region in more detail. 

 \vspace{1ex}
{\it (a) SPT entanglement  and  the topological Hamiltonian.--}
 \vspace{0.5ex}

\noindent 
Let us start by focusing on the  entanglement of the ground-state, which can  defined via the block reduced density matrices $\rho_\ell={\rm Tr}_{L-\ell}\{\ket{\epsilon_{\rm gs}}\bra{\epsilon_{\rm gs}}\}$. Among the various existing measures of entanglement on a bi-partite scenario~\cite{entanglement_measures}, the {\it entanglement entropies} (EEs) enjoy a privileged status within the realm of quantum many-body systems~\cite{ent_many_body,EE_many_body}. In particular, the so-called R\'enyi EEs  are defined for all $n>0$
\beq
\label{eq:REEs}
S_n(\ell,L)=\frac{1}{1-n}\log\big({\rm Tr}\left\{\left(\rho_\ell\right)^n\right\}\big),
\eeq 
and  include the Von Neuman entanglement entropy (VNEE)  $S(\ell,L)=-{\rm Tr}\{\rho_{\ell}\log(\rho_{\ell})\}=\lim_{n\to 1}S_n(\ell,L)$ as a limiting case.  Moreover, the R\'enyi EEs are related to the moments of the reduced density matrix~\cite{moments_es}, and can  be used to construct the full {\it entanglement spectrum} ~\cite{entanglement_spectrum}, which is defined by the set of all eigenvalues ${\rm ES}(\rho_\ell)=\sigma(H_{\rm E})$ of the entanglement Hamiltonian $H_{\rm E}$, obtained by expressing the reduced density matrix as an equilibrium Gibbs state $\rho_\ell\propto \ee^{-H_{\rm E}}$. 
 
For non-interacting systems, $H_{\rm E}$ has a closed-form expression as a quadratic operator that can be derived exactly from the exact diagonalization of the  original Hamiltonian~\cite{red_density_matrices_free}. For interacting systems, the entanglement spectrum $\epsilon_n\in {\rm ES}(\rho_\ell)$ can be obtained  approximately by the  MPS 
simulations~\cite{mps}, which give direct access to the Schmidt values $\lambda_n$ of the bipartition, such that $\epsilon_n=-2\log\lambda_n$.  As shown in~\cite{ES_spt}, the entanglement spectrum can be used to characterize SPT phases such as correlated topological insulators, as it displays an exact degeneracy  related to the existence of many-body edge modes ~\cite{ES_spt}. In Fig.~\ref{fig_ES_comparison}, we show how this feature is fulfilled clearly for the  correlated $\mathsf{AIII}$ topological insulator~\eqref{eq:CH_ham}. As a comparison, we  display the entanglement spectrum of the full Creutz-Hubbard ladder~\eqref{eq:CH_ham}, and that of the  RG-corrected topological Hamiltonian~\eqref{eq:single_part_creutz}, which 
can be calculated exactly via the two-point correlation functions~\cite{peschel_rho_red_corr}. Both methods display the aforementioned degeneracies for the correlated topological insulator~\eqref{eq:top_inv_CH_m0_int}, and show a clear quantitative  agreement providing an additional test of the validity of the RG-corrected topological Hamiltonian approach. 

It is also worth mentioning that our predictions can be provided experimentally. 
In ~\cite{dalmonte_papers} the authors proposed an immediate, scalable recipe for implementation of the entanglement
Hamiltonian, and measurement of the corresponding entanglement spectrum as spectroscopy of the
Bisognano-Wichmann Hamiltonian ~\cite{bisignano_paper} with synthetic quantum systems.

\vspace{1ex}
{\it (b) Bi-partite fluctuations and the Luttinger parameter.--}
\vspace{0.5ex}
 
We now explore   the scaling region of the quantum phase transition between the topological and trivial band insulators (see Fig.~\ref{fig_phase_diagram_zoom}) from the perspective of  bi-partite correlations. In Fig.~\ref{fig_REEs_comparison}, we represent the critical R\'enyi EEs~\eqref{eq:REEs} obtained by the two approaches as a function of the so-called chord length $D(\ell,L)=(2L/\pi)\sin(\pi\ell/L)$. As can be observed in this figure, the agreement between  effective RG model and full one, is very good for  weak interactions, but small deviations become apparent in the higher-order R\'enyi EEs as the Hubbard interactions increase and we move along the critical line of Fig.~\ref{fig_phase_diagram_zoom}. 

We would like to understand the origin of these small differences. On the one hand, they might simply be caused by inaccuracies in the RG-corrected imbalance~\eqref{eq:RG_imbalance}. However, for the range of Hubbard interactions hereby explored, we have found that the differences  in the critical points are vanishingly small. On the other hand, they might be caused by the approximations inherent to the topological Hamiltonian~\eqref{top_ham}, namely considering only the static part of the self-energy to define an effective single-particle Hamiltonian. In the present context, the critical properties of the topological Hamiltonian~\eqref{eq:single_part_creutz} are always governed by a single massless Dirac fermion, which might not be an accurate description of the scaling region of the full Creutz-Hubbard ladder as one moves along the critical line by increasing the interactions.

In order to explore this question further, we will explore the particular scaling of the critical bi-partite correlations with the chord distance. Due to non-extensive nature of the EEs $S_n(\ell,L)=S_n(L-\ell,L)$,  the bipartite correlations might be expected to be contained within the boundary of the bipartition, which leads to the so-called entropic logarithmic area laws~\cite{area_law_rmp}. As occurs for the VNEE~\cite{vn_entropy}, the R\'enyi EEs also display area-law violations   for one-dimensional critical systems~\cite{cft_entropy}, which contain information about the underlying conformal field theory (CFT) that describes the long-wavelength properties of the system~\cite{cft_book}. For open boundary conditions,   one finds that the R\'enyi EEs can be expressed as
\beq
S_n(\ell,L)=S_n^{\rm CFT}(\ell,L)+\delta S_n(\ell,L),
\eeq
which contains a logarithmic violation of the area law that depends on the central charge $c$ of the CFT
\beq
\label{eq:cft_renyi}
 S_n^{\rm CFT}(\ell,L)=\frac{c}{12}\left(1+\frac{1}{n}\right)\log D(\ell,L),
\eeq
together with a non-universal correction  $\delta S_n(\ell,L)$. We note that the logarithmic scaling of these EEs for various interactions displayed in Fig.~\ref{fig_REEs_comparison} is consistent with a central charge $c\approx1$, which  corresponds to the CFT of a free compactified boson~\cite{cft_book}. This agrees with the  leading-order scaling of the R\'enyi EEs  for the topological Hamiltonian~\eqref{eq:single_part_creutz}, such that the aforementioned deviations must be contained in the non-universal terms $\delta S_n(\ell,L)$. In particular, one possibility is that the Luttinger parameter, which is always fixed to $K=1$ for the RG-corrected topological Hamiltonian~\eqref{eq:single_part_creutz}, becomes  $K\neq1$ for the full Creutz-Hubbard ladder~\eqref{eq:CH_ham}. 

As  first realized  for the  Von Neumann EE with open boundary conditions~\cite{ee_corrections_luttinger}, the non-universal corrections to the R\'enyi EEs~\cite{renyis_corrections}   for  systems whose critical behavior is described by the free-boson CFT (i.e. Luttinger liquids~\cite{lutt_liq}) find the following closed-form expression 
\beq
\label{eq:corrections_renyis}
\delta S_n(\ell,L)=F_n\left(\frac{\ell}{L}\right)\frac{\cos(2k_{\rm F}\ell+\omega)}{\big(2\sin( k_{\rm F})D(\ell,L)\big)^{K/n}},
\eeq
where $k_{\rm F}$ is Fermi's momentum, $K$ is the so-called Luttinger parameter, $F_n(\ell/L)$ is a scaling function, and $\omega$ is a constant phase-shift that controls the oscillating nature of the corrections. For instance, for spin-$\half $ Heisenberg-type Luttinger liquids (i.e. XXZ model), where $k_{\rm F}=\pi/2$, one finds  $\omega=0$ such that the correlations display a characteristic alternating behavior~\cite{ee_corrections_luttinger,renyis_corrections}. On the other hand,   for  hard-core bosons with dipolar interactions at quarter filling, where $k_{\rm F}=\pi/4$, one needs to set $\omega\approx\pi/2$ to capture the  oscillations~\cite{REE_K_dipolar_boson}.

We note that this characteristic oscillatory behavior is in principle useful to extract numerically the corresponding Luttinger parameter~\cite{taddia_thesis}. Unfortunately, for the Creutz-Hubbard ladder, we have found that these oscillations are absent, such that the fitting procedure to extract the Luttinger parameter is not sufficiently accurate.
We now provide an alternative route to extract the Luttinger parameter by exploring the quantum noise in models with a conserved quantity, such as the total  fermion number $N=\sum_{j}\big(c_{j,{\rm u}}^{\dagger}c_{j,{\rm u}}^{\phantom{\dagger}}+c_{j,{\rm d}}^{\dagger}c_{j,{\rm d}}^{\phantom{\dagger}}\big)$ in the Creutz-Hubbard ladder~\eqref{eq:CH_ham}. As first realized in~\cite{fluctuations_entropy},  noise in certain quantum-transport scenarios  can be directly related to entropic entanglement measures. In fact, for free-fermion systems~\cite{noise_cumulants_entropy_free_fermions}, it can be shown rigorously that any R\'enyi EE~\eqref{eq:REEs} can be reconstructed from the knowledge of the noise cumulants for a certain bi-partition of the  system. In free-fermion systems where the noise is purely Gaussian, it suffices to consider the second cumulant (i.e. variance) of the conserved quantity restricted to the bi-partition $N_\ell$, which is sometimes referred to as the bi-partite fluctuations
\beq
S^{\rm free}_{n}(\ell,L)=\frac{\pi^2}{12}\!\left(1+\frac{1}{n}\right)\!\mathcal{F}(\ell,L),\hspace{1.5ex}\mathcal{F}(\ell,L)=\langle(N_\ell-\langle N_\ell\rangle)^2\rangle.
\eeq

In the case of models that can be mapped onto the free-boson $c=1$ CFT (i.e. Luttinger liquids), a similar expression holds as   fluctuations are Gaussian~\cite{noise_cumulants_lutt}. The  difference is that the Luttinger parameter also appears in the proportionality  
\beq
\label{eq:LL_fluctuations}
S^{\rm CFT}_{n}(\ell,L)=\frac{\pi^2}{12K}\left(1+\frac{1}{n}\right)\mathcal{F}(\ell,L).
\eeq
Using the expression of the CFT prediction for the Renyi-entropy~\eqref{eq:cft_renyi}, one finds that the bi-partite fluctuations of a Luttinger liquid with conserved particle number should scale with the chord distance as follows
\beq
\label{eq:scaling_bip_fluctuations}
\mathcal{F}(\ell,L)=\frac{K}{\pi^2}\log(D(\ell,L)).
\eeq
Accordingly, the bi-partite fluctuations of a critical Luttinger liquid also show an area-law violation, and one can use them to extract the underlying Luttinger parameter $K$. We note that this additional conservation  law  could also be combined with the entanglement entropies, leading to an equipartion that may allow for more accurate extractions of the $K$ parameter~\cite{sierra_paper}.

    \begin{figure}
\centering
\includegraphics[width=0.9\columnwidth]{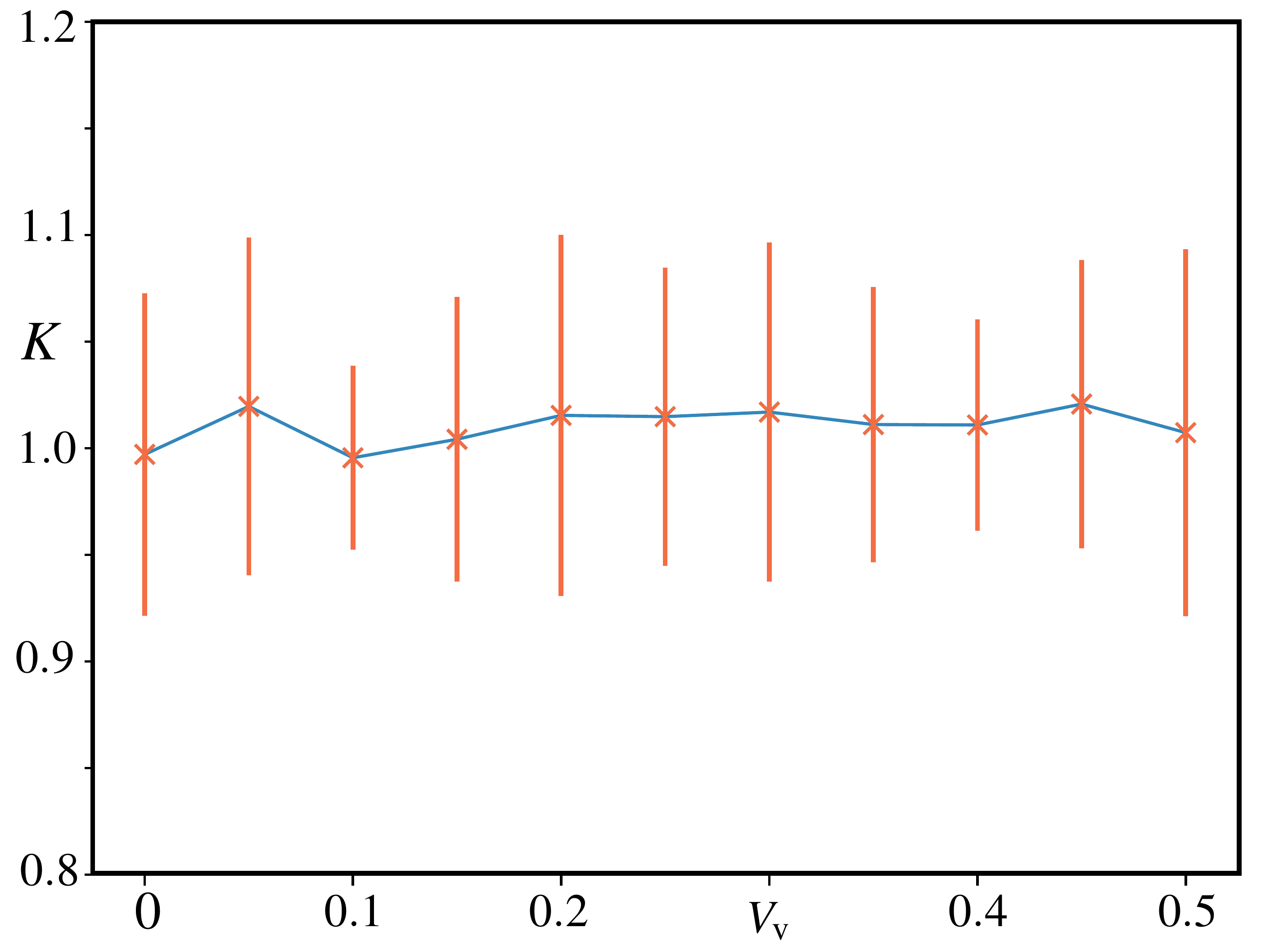}
\caption{ {\bf  Luttinger parameter at criticality: } The Luttinger parameter $K$ is extracted by fitting the bi-partite fluctuations $\mathcal{F}(\ell,L)$ obtained by the MPS numerical simulations to the linear logarithmic scaling with the chord length of  Eq.~\eqref{eq:scaling_bip_fluctuations}. For each critical point $(\Delta\epsilon, V_{\rm v})$, we extract the thermodynamic value of the Luttinger parameter by extracting the corresponding $K_L$ values for various lengths $L/a\in[64,192]$, and fitting them to $K_L=K+c_1/L+c_2/L^2$.}
\label{fig_Luttinger}
\end{figure}

In Fig.~\ref{fig_Luttinger}, we represent the Luttinger parameter $K$ for the critical Creutz-Hubbard ladder in the thermodynamic limit. We modify the Hubbard interaction strengths $V_{\rm v}$, and fix the energy imbalance $\Delta\epsilon$ to the RG-predicted value $\Delta\epsilon_{\rm c}$ of Eq.~\eqref{eq:two_loop_critical_line}, such that the system lies at the critical line. The bi-partite fluctuations  $\mathcal{F}(\ell,L)$ are numerically calculated using our MPS algorithm, and the value of $K$ is extracted following the procedure detailed in the corresponding caption. As this figure shows, the Luttinger parameter remains very close to the value $K\approx 1$, which would correspond to a free massless Dirac fermion as the CFT controlling the critical scaling. Let us note, however, that this numerical procedure has turned out to be  very sensitive to small variations of the parameters. The error bars displayed in Fig.~\ref{fig_Luttinger} correspond to the Luttinger parameter obtained by modifying the energy imbalance within $(\Delta\epsilon_{\rm c}- 0.01t,\Delta\epsilon_{\rm c}+ 0.01t)$ for 10  different values of $\Delta\epsilon$, and obtaining the standard deviation of all the corresponding Luttinger parameters $K$. Accordingly, although it seems that the Luttinger parameter remains at $K\approx1$ as the interactions are increased, our numerical results are not conclusive, as minor modifications in $\Delta\epsilon$ can lead to large variations in $K$. In case $K=1$ all along the critical line, the aforementioned differences might be caused by further non-universal and finite-size corrections. However, we note again that due to the sensitivity found in this problem, further numerical analysis
shall be required in the future to clarify this point.
 
\section{Conclusions and Outlook}
\label{sec:conclusions}

In this work, we have presented a detailed study of the use of the RG to obtain an  effective relativistic QFT that provides a continuum long-wavelength description of correlated topological insulators. Our description is valid for a variety of minimal models that serve as representatives of various topological insulators~\cite{Bernevig}, and are related to self-interacting fermionic lattice field theories in various dimensions~\cite{lattice_book}, which we refer to as topological Wilson-Hubbard matter. 

We have shown that a Wilsonian RG, where the effect of integrating high-energy modes as the cutoff is reduced to focus on long-wavelength phenomena leads to a renormalization of the microscopic couplings, provides a neat description of these correlated topological insulators. At the so-called tree level, and generic to various dimensions, the RG approach offers a neat qualitative connection between the topological invariants in the long-wavelength limit and the flat-band limit of topological insulators. This connection becomes quantitative by using the RG flows of the parameters in connection to the so-called topological Hamiltonian, which includes static self-energy corrections to characterize the many-body topological invariant of the correlated topological insulator. 

Going beyond tree level, we have shown that a loop expansion of the Wilsonian RG offers a quantitative route to understand the topological phase transitions that occur in Wilson-Hubbard matter, separating the correlated topological insulator from other non-topological phases. We have benchmarked the  two-loop RG predictions for a particular 1D model, the imbalanced Creutz-Hubbard ladder, to quasi-exact numerical simulations based on matrix product states. This numerical comparison shows a very good agreement in the determination of the critical line determining the topological quantum phase transition, as well as various bi-partite correlations of the topological phase and the critical line. This benchmark may motivate the extension of the RG calculations to higher-dimensional correlated topological insulators in the future.

Let us now comment on the interpretation of our results from the perspective of dynamical mass generation in QFTs. As advanced in the main text, 4-Fermi continuum QFTs can display chiral symmetry breaking via dynamical mass generation~\cite{GN_model}. However, the scaling of the dynamically-generated mass with the 4-Fermi interaction strength is non-perturbative, which differs from the numerical matrix-product-state simulations for the particular lattice model studied in this work. Therefore, we believe that it is not possible to capture the physics of this type of correlated topological insulators by a continuum QFT if one insists on disregarding the heavy Wilson fermions lying at the cutoff of the lattice field theory, and  considers only the  4-Fermi continuum QFTs for the light Wilson fermions. Our RG calculations indicate that the role of these heavy fermions in the RG flows of  Wilson-Hubbard topological matter is very important. 

In this context, the mapping of the imbalanced Creutz-Hubbard ladder~\cite{GNW_model_spt}, and other 1D correlated topological insulators~\cite{kuno_SSH_int}, to  discretized Gross-Neveu lattice field theories offers a more concrete perspective on such a dynamical mass generation. Large-$N$ calculations show that, on top of the dynamically-generated mass generation~\cite{GN_model}, there are additive renormalizations of the mass when one considers the complete lattice action~\cite{GNW_model_spt,kuno_SSH_int}. From the perspective of the continuum limit and our RG calculations, it seems to us that the interplay of light and heavy Wilson fermions in the RG flows is actually capturing this additive mass renormalization, and the two-loop calculations allow us to go beyond the $N\to\infty$ predictions for weak interactions~\cite{GNW_model_spt,kuno_SSH_int}. It would be very interesting to explore this connection further, and see if both approaches can  even be combined to provide a more accurate description of the topological phase transition, especially in higher dimensions.

Let us finally also comment on recent efforts to formulate a different RG for correlated topological insulators, which offers a very interesting alternative to the standard Wilsonian scheme~\cite{rg_correlated_top_insulators}. In this approach, one can build a scaling theory from the fact that the topological curvature, e.g. Berry connection, associated to the topological invariant presents some divergence at a topological quantum phase transition. Instead of changing the cutoff and integrating fast modes {\it a la} Wilson, this approach obtains RG flows by studying how this curvature changes with a flow of the microscopic parameters around high-symmetry points in momentum space~\cite{curvature_rg}. It would be very interesting to compare the predictions of this so-called curvature renormalization group (CRG) to our more-standard Wilsonian RG, in particular in the context of the imbalanced Creutz-Hubbard ladder, where the quasi-exact numerical results allow for a very detailed benchmark.

Finally, let us comment on one of the points raised in the introduction, namely the possibility of using condensed-matter analogues to explore models of high-energy physics. Besides the solid-state materials mentioned, ultracold gases of neutral atoms trapped in periodic light potentials~\cite{cold_atom_review} have become a very-flexible platform to explore a variety of analogues that can be described accurately by the particular model under study, acting as quantum simulators~\cite{q_sim_review_1}. We note that the Wilson-Hubbard topological matter described in this work may find experimental realizations along the lines of~\cite{wilson_atoms,GNW_model_spt,kuno_SSH_int,wilson_hauke,wilson_kuno,KHS18} in the field of cold atoms.

\section*{Acknowledgments}

We acknowledge interesting discussions with S. Hands. A.B. acknowledges support from Spanish MINECO Projects FIS2015-70856-P,  and CAM PRICYT project  QUITEMAD+ S2013/ICE-2801.
\noindent
E.T. and M.R. acknowledge computational time from the Mogon cluster of the JGU (made available by the CSM and AHRP), S. Montangero for a long-standing collaboration on the flexible Abelian Symmetric Tensor Networks Library employed here, as well as J. J\"{u}nemann for his participation in early stages of this work. M.R. acknowledges also support by the DFG under Project RI 2345/2-1.
\noindent
M.L. and E.T. acknowledge the Spanish Ministry MINECO (National Plan
15 Grant: FISICATEAMO No. FIS2016-79508-P, SEVERO OCHOA No. SEV-2015-0522, FPI), European Social Fund, Fundació Cellex, Generalitat de Catalunya (AGAUR Grant No. 2017 SGR 1341 and CERCA/Program), ERC AdG OSYRIS, EU FETPRO QUIC, and the National Science Centre, Poland-Symfonia Grant No. 2016/20/W/ST4/00314. 
\noindent
G.S. acknowledges support from grant IFT Centro de Excelencia Severo Ochoa SEV-2016- 0597, 
Grant No. FIS2015-69167-C2-1-P from the Spanish government, and QUITEMAD+ S2013/ICE-2801
from Comunidad Autonoma Madrid.


\end{document}